\documentclass[%
 aip,
 amsmath,amssymb,
 reprint,%
]{revtex4-1}

\usepackage{graphicx}
\usepackage{dcolumn}
\usepackage{bm}

\usepackage[utf8]{inputenc}
\usepackage[T1]{fontenc}
\usepackage{mathptmx}
\usepackage{etoolbox}
\usepackage{pstricks}

\graphicspath{{figures/}}

\makeatletter
\def\@email#1#2{%
 \endgroup
 \patchcmd{\titleblock@produce}
  {\frontmatter@RRAPformat}
  {\frontmatter@RRAPformat{\produce@RRAP{*#1\href{mailto:#2}{#2}}}\frontmatter@RRAPformat}
  {}{}
}%
\makeatother

\newrgbcolor{dred}{.67 0 0}

\newcommand{\bA}{{\bf A}}

\newcommand{\bv}{{\bf v}}
\newcommand{\bp}{{\bf p}}
\newcommand{\bP}{{\bf P}}
\newcommand{\bJ}{{\bf J}}
\newcommand{\br}{{\bf r}}
\newcommand{\ba}{{\bf a}}
\newcommand{\bk}{{\bf k}}

\newcommand{\ph}{{\text{(ph)}}}
\newcommand{\tb}{\text{b}}

\newcommand{\bE}{{\bf E}}
\newcommand{\bH}{{\bf H}}

\begin{document}

\preprint{AIP/123-QED}

\title[]{The Schr\"{o}dinger equation in the complex plane and quantum entanglement}

\author{Vassiliy Lubchenko*}%
 \email{vas@uh.edu}
\affiliation{Department of Chemistry, University of
  Houston, Houston, TX 77204-5003}%
\affiliation{Department of Physics, University of Houston, Houston, TX
  77204-5005} \affiliation{Texas Center for Superconductivity,
  University of Houston, Houston, TX 77204-5002}
  
\date{\today}

\begin{abstract}

We formulate a continuity equation for the Schr\"odinger equation in
the complex space. We define a complex momentum by normalizing the
complex current by the particle density. This momentum is a quantum
analog of the classical, kinematic momentum analytically continued
into the complex plane. The kinematic momentum and the gradient of the
wavefunction's phase each represent a fluid-like flow in the complex
plane; the phase-gradient flow is incompressible. The zeros of the
wavefunction give rise to simple poles in the momentum. The poles
manifest as irrotational vortexes in the phase-gradient flow, while
critical points of the wavefunction present as rigid body-like
rotational flows of the kinematic momentum. The discrete nature of
elementary excitations comes about inherently because the quantity of
the poles is automatically integer. An exact quantization condition is
subsequently formulated, which reduces to the Bohr-Sommerfeld
condition in the semiclassical limit. We establish a priori that the
Bohr-Sommerfeld condition must be exact for the Harmonic
Oscillator. We show that the kinetic energy is a sum of contributions
of the average value and fluctuations, respectively, of the kinematic
momentum. The zero-point vibrations within bound states are solely due
to the fluctuations of the momentum and manifest as rigid-body flows
at infinity. The momentum poles---and hence the wavefunction's
zeros---can be viewed as emergent, consistent with the remarkable
property of quantum entanglement exhibited by standing wave solutions
of the Schr\"odinger equation.

\end{abstract}

\maketitle

\section{\label{intro} Introduction}

The Schr\"{o}dinger equation is a differential equation that retains
its form when analytically continued into the complex plane. This
continuous, local symmetry becomes particularly manifest at the
semiclassical level,~\cite{Kemble1937, LLquantum} whereby one has a
considerable freedom to deform integration contours in the complex
plane without affecting the integral's value. One may expect that this
symmetry is associated with a conservation law, similarly to how gauge
invariance in electrodynamics is intrinsically connected to charge
conservation.~\cite{LLclass, itzykson2012quantum, Peskin} And indeed,
already Riemann~\cite{klein1893riemann} points out that
complex-differentiability of a function implies there is an associated
incompressible flow in the complex plane.

Motivated in part by those notions, one may ask whether a consistent
quantum mechanical description can be formulated that allows for
transport in the complex plane at the outset: On the one hand, the
momentum operator essentially generates translations in space. On the
other hand, already formulations of the momentum in terms of
real-valued derivatives establish that classically forbidden regions
must be accessible by matter. Despite singularities that arise near
the classical turning points,~\cite{doi:10.1073/pnas.14.2.178,
  PhysRev.48.549, 10.1063/1.1705112, 10.1063/1.1677086,
  BerryMount1972} semiclassical analyses indicate that when continued
analytically into classically forbidden regions, momentum becomes
complex-valued. But so should displacements it generates! Conversely,
if one were to presume that movement of inertial matter can no longer
be connected to momentum as one crosses over to classically forbidden
regions, one would have to explicitly work out how this presumed
change comes about.  Of great interest in its own right, mass
transport in classically forbidden regions displays puzzling features,
such as seemingly superluminal tunneling
velocities.~\cite{WINFUL20061, PhysRevLett.91.260401, BARTON1986322,
  RevModPhys.66.217}

Superluminal-like effects are not limited to classically-forbidden
regions, either. Standing-wave solutions of the Schr\"odinger equation
can exhibit zeros, implying the corresponding locations would have to
be strictly avoided by matter. This was of grave concern to early
practitioners of Quantum Mechanics, to quote
Kemble's~\cite{Kemble1937} Chapter III(19d): ``But we are in apparent
difficulty if we ask how it happens that sometimes we find a particle
on one side of the node and sometimes on the other---never right on
the node itself. For if we think of particles vibrating back and
forth, unless they attain an infinite velocity in passing through the
nodes, they must spend some time in the neighborhood of each.''  The
great difficulty of addressing that early concern from a purely
corpuscular standpoint can be turned around to question the utility of
the corpuscular standpoint itself. Yet the apparent property of matter
to be at least somewhat delocalized does not obviate that it takes
time to accelerate it. If anything, a (putative) particle's not being
in one place makes the task of accelerating it toward a certain
location only harder.

Once established, standing-wave configurations can enable one to use
local information to know the system's behavior very far away---with
considerable confidence and without delay. To do so, one needs to
determine the shape of the signal, but only locally, and then continue
the shape in some regular fashion; in the limit of vanishing external
potential one would simply have to continue the signal
periodically. Standing waves thus present a simple yet striking
realization of the important phenomenon of quantum
entanglement.~\cite{10.1119/1.5003808} This seeming
action-at-a-distance property of standing waves and, generally, any
coherent mixtures of propagating waves, results from a symmetry
breaking: In such a mixture, the phase of an individual wave is no
longer arbitrary, but, instead, becomes pegged to the phases of the
other waves, thus breaking spatial symmetry. Whether the wavefunction
strictly vanishes or remains finite at the troughs of the density
profile, the latter profile is no longer spatially uniform but,
instead, becomes locally modulated in a regular fashion, thereby
revealing the wave-like nature of the underlying signal. The length
scale of the local modulation varies relatively slowly in space and is
directly related to the wavelength of the constituent waves.

That standing waves imply long range correlations invariably raises
interesting mechanistic questions and is not unique to
quantum-mechanical setups, of course. For instance, a solid---quantum
or classical---embodies a standing matter wave that has a
non-vanishing sheer modulus, thus allowing one to transfer momentum
over macroscopic lengths, even as individual particles only move
within a small fraction of the particle spacing. This type of standing
matter wave is a broken-ergodicity state that emerges following a
discontinuous transition from a uniform liquid.~\cite{L_AP, LW_ARPC,
  DGL, LTLS2026arxiv} Another example is the metal-insulator
transition via the formation of a standing charge-density wave, as
caused by a continuous symmetry breaking; the phase relation between
the constituent electronic waves determines the insulator's
type.~\cite{LKgap}

In the formalism presented here, the mass current and the associated
momentum~\cite{Messiah, Berry2013} are made complex-valued at the
onset. The resulting description makes explicit those seemingly
superluminal features of solutions of the Schr\"odinger equation
discussed above. The formulation comes along with a rather vivid
visual representation of several quantum mechanical phenomena of
practical interest: bound states, tunneling, scattering in general,
and zero-point oscillations. In this representation, the complex space
is fully covered by non-intersecting trajectories generated using the
complex momentum. The complex momentum is well behaved almost
everywhere in the complex plane except in the vicinity of the
wavefunction's zeros.  At the zeros, the momentum diverges by way of
simple poles---c.f. the Kemble quote above. Alongside, a separate
momentum-like quantity exhibits irrotational vortexes; the latter
momentum is associated with the spatial variation of the
wavefunction's phase. The zeros of the wavefunction and the associated
vortexes are arranged along strings of finite or infinite length that
separate regions of smooth, classical-like flows; the spacing between
adjacent vortexes within a string is regular. Standing waves and,
generally, interference patterns result from strings of poles that run
in the vicinity of the real axis. Although the complex momentum
diverges right at the axis of each pole, the actual mass currents, if
any, remain finite because the wavefunction vanishes concurrently. The
smooth behavior of the actual particle flux along the real axis can be
viewed, then, as a cancellation of two singularities.

The complex momentum associated with the complex current obeys a
Riccati equation that is locally equivalent to the underlying
Schr\"odinger equation. Yet, if one were to adopt, at the onset, the
Riccati equation as the equation of motion---while assuming that the
equation be valid everywhere in the complex plane---physical
configurations that involve zeros in the wavefunction would not be
found. Instead, the momentum poles must enter the description as an
externally added construct. The putative locations of individual poles
must be excluded from the domain of the validity of the equations of
motion. In this way, the momentum poles are similar to point-like
defects in an otherwise smooth continuum. The residues at the poles
can be found self-consistently already from the Riccati equation. The
locations of the poles are found self-consistently, too, yet to obtain
the pertinent equation---i.e. the Schr\"odinger equation itself---one
must first multiply the Riccati equation by the wavefunction, and then
allow the wavefunction to vanish in isolated points. This enables
pole-containing solutions by effectively regularizing them.

From this perspective, standing-wave solutions of the Schr\"odinger
equation can be viewed---much like elastic solids---as an emergent
state that originates from an instability toward a symmetry-lowered
state. Similarly to other cases of symmetry breaking, the relative
phases of the constituent propagating waves---and hence the precise
locations of the crests and troughs of the standing wave---are
determined by the external potential, whose spatial variation can be
vanishingly weak. Exact quantization conditions can be formulated that
boil down to counting the momentum poles and can be thought of as a
postmortem analysis of the wavefunction that had resulted from an
instability-induced transition. In this sense, elementary excitations
can be thought of as adding an integer number of ``defects'' in the
complex plane. The latter notion drives home that elementary
excitations are inherently discrete. We will also see directly that to
ensure normalizability of the wavefunction, the number of added
defects must be finite.

The article is organized as follows: In Section~\ref{flux}, we extend
the conventional continuity equation associated with the Schr\"odinger
equation into the respective complex planes for the spatiotemporal
coordinates. The pertinent current is a complex-valued quantity whose
real part, when computed on the real axis, is the conventional
particle current~\cite{Kemble1937, LLquantum, Sakurai1993Modern}
associated with the Schr\"odinger equation on the real axis. External
potentials enter the complex continuity equation in the form of
particle sinks/sources. A variety of momenta can be ascribed to the
wavefunction.  One type is obtained by normalizing the current by the
particle density and is a quantum analog of the classical kinematic
momentum. This momentum satisfies a Riccati equation associated with
the stationary Schr\"odinger equation.

In Section~\ref{continuum}, we take a continuum-mechanics perspective
on the Schr\"odinger equation, largely informed by Riemann's idea of
incompressible flows in the complex plane. We find that the complex
conjugate of the quantum canonical momentum represents such an
incompressible flow. The latter flow is directed along the lines of
constant particle density and, at the same time, orthogonally to the
lines of constant phase of the wavefunction.  We co-opt the complex
formulation of two-dimensional continuum mechanics to characterize the
spatial distribution of the wavefunction's zeros and critical points
by evaluating certain closed-loop integrals of the kinematic and
canonical momenta, respectively, similarly to how one counts and
characterizes vortexes in hydrodynamics. We thus obtain a quantization
condition that is exact. If applied to the semiclassical wavefunction,
it yields the venerable Bohr-Sommerfeld formula. There, we also find
that zero-point vibrations can be alternatively thought of as a result
of breaking the rotational symmetry for momentum flows at infinity.

In Section~\ref{flows}, we use the complex-momentum representation to
illustrate several important quantum-mechanical phenomena, including
scattering and tunneling, from the continuum-mechanics perspective
afforded by the momentum flows. A scattering potential serves to
deflect the flows toward the imaginary axis. So much so that at
energies below the barrier's col, the arrived trajectories for the
transmitted signal originate on the {\em outgoing} side of the
barrier.

In Section~\ref{fluct}, we establish an effective distribution for the
complex momentum and find that the kinetic energy can be presented as
a sum of contributions due to the average value and fluctuations,
respectively, of the kinematic momentum. The zero-point oscillations
come about as originating exclusively from the latter fluctuations;
these represent an intrinsic, defect-less background, in the complex
plane, on top of which momentum poles are added.  There we also see
that the fluctuation term is the sole cause of phase shifts during
scattering, while showing that the exactness of the Bohr-Sommerfeld
condition for the Harmonic Oscillator is rooted in the properties of
the oscillator's momentum in the complex
plane. Section~\ref{conclusions} provides a summary and discussion.

\section{Complex continuity equations, mass current, and momentum}
\label{flux}

Throughout, we use labels ``1'' and ``2'' to denote, respectively, the real and imaginary part of a complex-valued quantity. For instance, one has for the temporal coordinate
\begin{align}
    t = t_1 + i \, t_2 \label{t} \\
    t^* = t_1 - i \, t_2 \label{t*}
\end{align}
and likewise for the spatial coordinates as well as their combinations
such as the gradient $\nabla$; $i=\sqrt{-1}$ is the imaginary
unit. Consider a complex-valued function $g(x_1, x_2) = g_1(x_1, x_2)
+ i g_2(x_1, x_2)$ that is differentiable in terms of its two
real-valued arguments $x_1$ and $x_2$, respectively. One can readily
convince oneself that the increment in $g$ incurred by independently
varying $x_1$ and $x_2$ can be expressed as a linear combination of
increments $dx = dx_1 + i \, dx_2$ and $dx^* = dx_1 - i \, dx_2$,
respectively, of the two complex-valued variables $x = x_1 + i \, x_2$
and $x^* = x_1 - i \, x_2$:
\begin{eqnarray} \label{gxx}
    &&g(x_1 + d x_1, \: x_2 + d x_2) - g(x_1, x_2) \\ &&=  \frac{1}{2} \left( \frac{\partial g}{\partial x_1} - i \frac{\partial g}{ \partial x_2} \right) dx + \frac{1}{2} \left( \frac{\partial g}{\partial x_1} + i \frac{\partial g}{\partial  x_2} \right) dx^* + o(|dx|). \nonumber
\end{eqnarray}
Consequently, we define the Wirtinger derivatives according
to:~\cite{Wirtinger1927, gunning1965analytic}
\begin{align} 
    \frac{\partial}{\partial x} &= \frac{1}{2} \left( \frac{\partial}{\partial x_1} - i \frac{\partial}{ \partial x_2} \right) = \frac{1}{2} \left( \frac{\partial}{\partial x_1} + \frac{\partial}{ \partial (i x_2)} \right) \label{dx} \\
    \frac{\partial}{\partial x^*} &= \frac{1}{2} \left( \frac{\partial}{\partial x_1} + i \frac{\partial}{\partial  x_2} \right) = \frac{1}{2} \left( \frac{\partial}{\partial x_1} - \frac{\partial}{\partial (i x_2)} \right), \label{dx*}
\end{align}
and likewise for the spatial gradient: $\nabla = (\nabla_1 -i
\nabla_2)/2$, $\nabla^* = (\nabla_1 + i \nabla_2)/2$. We will
occasionally use the prime as a shorthand for differentiation with
respect to a complex spatial variable, but not its complex conjugate:
$f' \equiv \partial f/\partial x$.

For a {\em complex}-differentiable function $f(x_1, x_2) = f(x_1 + i x_2)$, the second term on the r.h.s. of Eq.~(\ref{gxx}) vanishes by construction, while the Cauchy-Riemann conditions can be compactly written as one, complex-valued equation:  
\begin{equation} \label{fx*}
\frac{\partial f}{\partial x^*} = \frac{1}{2} \left[ \frac{\partial f(x_1 + i x_2)}{\partial x_1} - \frac{ \partial f(x_1 + i x_2)}{\partial (i x_2)} \right]=  0.
\end{equation}
or, equivalently,
\begin{equation} \label{f*x}
    \frac{\partial f^*}{\partial x} =0.
\end{equation}
Since we allow for $f$ to be complex-valued on the real axis, $f^* \equiv [f(x)]^* \ne f(x^*)$ generally, but Eqs.~(\ref{fx*}) and (\ref{f*x}) still hold. 

Transitioning from complex differentiation to real-valued differentiation and back is done by inverting Eqs.~(\ref{dx})-(\ref{dx*}) 
\begin{align} \label{dx1}
    \frac{\partial}{\partial x_1} &= \frac{\partial}{\partial x} + \frac{\partial}{\partial x^*} \\
    \frac{\partial}{\partial x_2} &= i \left( \frac{\partial}{\partial x} - \frac{\partial}{\partial x^*} \right) \label{dx2}
\end{align}
and using Eqs.~(\ref{fx*})-(\ref{f*x}). This yields
\begin{align} \label{dfx1}
    \frac{\partial f}{\partial x_1} &= \frac{\partial f}{\partial x} \\ \label{df*x1}
    \frac{\partial f^*}{\partial x_1} &= \frac{\partial f^*}{\partial x^*}
\end{align}
etc. The Wirtinger derivatives are particularly convenient, when one
needs to vary objects that are products of functions of a complex
variable and its complex conjugate, respectively. Consider, for
instance, the following object:
\begin{equation} \label{object}
    \frac{1}{i} \psi^* \nabla_1 \psi = \frac{1}{i} \psi^* \nabla \psi = \frac{1}{i} \nabla \psi^* \psi = \frac{1}{i} \nabla |\psi |^2,
\end{equation}
where $\psi$ is a complex-differentiable function and we used
Eq.~(\ref{dfx1}) to obtain the second equality, Eq.~(\ref{f*x}) to
obtain the third equality. By Eq.~(\ref{dx}) one has, then, for the
real and imaginary part, respectively, of this object:
\begin{equation} \label{Rpnp}
    \Re \frac{1}{i} \psi^* \nabla_1 \psi = - \frac{1}{2} \nabla_2 |\psi|^2,
\end{equation}
\begin{equation} \label{Ipnp}
    \Im \frac{1}{i} \psi^* \nabla_1 \psi = - \frac{1}{2} \nabla_1 |\psi|^2,
\end{equation}
since the quantity $|\psi|^2$ is real-valued. At the same time, we recognize that the real part
\begin{eqnarray}
    \Re \frac{1}{i} \psi^* \nabla_1 \psi &=& \frac{1}{2 i} (\psi^* \nabla_1 \psi - \psi \nabla_1 \psi^*)
\end{eqnarray}
gives, up to a multiplicative constant, the conventional particle
current corresponding to a wavefunction $\psi$, in the absence of the
vector potential.~\cite{LLquantum, Sakurai1993Modern} Thus we
tentatively observe that particle flux along the real axis can be
evaluated by computing the gradient of the density $|\psi|^2$ along
the {\em imaginary} axis; one need not know the wavefunction's phase,
c.~f. Chapter VI \S~4 of Ref.~\cite{Messiah} or Eq.~(19.4a) of
Ref.~\cite{LLquantum} At the same time, we notice that the current
along the real axis has a counterpart directed along the imaginary
axis in the complex plane; this counterpart is determined,
symmetrically, by the gradient of the density along the real axis.

Consider now the Schr\"odinger equation
\begin{equation} \label{pt}
    i \hbar \frac{\partial \psi}{\partial t_1} = \hat H \psi,
\end{equation}
where the Hamiltonian $\hat H$ is 
\begin{equation} \label{H}
    \hat H = \frac{1}{2m}\left( -i \hbar \nabla_1 - \bA \right)^2 + V,
\end{equation}
as appropriate for a non-relativistic, spinless particle subjected to
an external electrostatic potential $V$ and vector potential $\bA$. We
have set the speed of light $c$ and particle's charge $q$ to unity for
typographical convenience; complete formulas are recovered by
replacing $V \to qV$, $\bA \to (q/c) \bA$. We limit ourselves to
potentials $V$ and $\bA$ that can be analytically continued off the
real axis onto a strip of non-vanishing width that contains the real
axis. Despite some degree of utility, this setup excludes important
cases, such as the Coulomb potential. Note the derivation allows for
$V$ and $\bA$ to be complex-valued already on the real axis.

Assuming the wavefunction is complex-differentiable,  one has, analogously to Eq.~(\ref{dfx1}):
\begin{equation} \label{tp1}
    \frac{\partial \psi}{\partial t} =  \frac{\partial \psi}{\partial t_1}. 
\end{equation} 
and
\begin{equation} \label{p2p}
    \left(\nabla_1 - \frac{i \bA}{\hbar}\right)^2 \! \psi =
    \left(\nabla - \frac{i \bA}{\hbar}\right)^2 \! \psi.
\end{equation}
Consequently, one can readily write down the Schr\"odinger equation
equivalently in terms of complex derivatives:
\begin{equation} \label{SchC}
    i \hbar \frac{\partial}{\partial t} \psi = \left[
      \frac{1}{2m}\left( -i \hbar \nabla - \bA \right)^2 + V \right]
    \psi.
\end{equation}
Eq.~(\ref{SchC}) explicitly demonstrates that the Schr\"odinger equation is invariant with respect to analytic continuation into the complex plane, again, subject to the constraint that $\psi$ be complex-differentiable in the vicinity of the real axis. 

Multiplying Eq.~(\ref{SchC}) by $\psi^*$ and using
\begin{equation} \label{tp2}
    \frac{\partial}{\partial t} |\psi|^2 =  \psi^* \frac{\partial \psi}{\partial t}, 
\end{equation} 
and
\begin{equation} \label{p2p2}
   \psi^* \!\left(\nabla - \frac{i \bA}{\hbar}\right)^2 \! \psi = \left(\nabla - \frac{i \bA}{\hbar}\right)^2 \! |\psi|^2
\end{equation}
yields a Schr\"odinger-like equation for the density:
\begin{equation} \label{Schdens}
     i \hbar \frac{\partial}{\partial t} |\psi|^2 = \left[ \frac{1}{2m}\left( -i \hbar \nabla - \bA \right)^2 + V \right] |\psi|^2.
\end{equation}
Eq.~(\ref{Schdens}) can be equivalently re-written as a continuity equation for the mass density 
\begin{equation}
    \rho = m |\psi|^2
\end{equation}
with particle sinks and sources:
\begin{equation} \label{contC}
    \left(\frac{\partial}{\partial t} + \frac{i \, V}{\hbar}\right) \rho = - \frac{1}{2} \left(\nabla - \frac{i \bA}{\hbar}\right) \bJ,
\end{equation}
where we define the complex momentum-current $\bJ$ according to:
\begin{equation} \label{jC}
    \bJ = \frac{\hbar}{i} \left(\nabla - \frac{i \bA}{\hbar}\right) |\psi|^2 = \psi^* \frac{\hbar}{i} \left(\nabla_1 - \frac{i \bA}{\hbar}\right) \psi.
\end{equation}
This definition of complex current is internally consistent: Implicit
in the complex formulation (\ref{SchC}) of the Schr\"odinger equation
is the complex extension of the momentum operator:
\begin{equation} \label{Pop}
    \hat \bP = \frac{\hbar}{i} \nabla = \frac{\hbar}{2 i} (\nabla_1 - i \nabla_2).   
\end{equation}
Analogously to how its real-valued counterpart generates translation along the real axis,~\cite{LLquantum} the complex-momentum operator generates translation in the complex plane:
\begin{equation}
    e^{i \ba \hat \bP/\hbar} \varphi(\br) = \varphi(\br + \ba), 
\end{equation}
if $\varphi$ is complex-differentiable, as can be seen by
Taylor-expanding the exponential. Note the definition (\ref{jC}) of
the complex current implies that
\begin{equation}
    \bJ = (\hat \bP - \bA) |\psi|^2.
\end{equation}

Analogously to how we do this for the real-valued
momentum,~\cite{Messiah, Berry2013} one may then associate with the
complex momentum current $\bJ$ a complex momentum $\bp$---not to be
confused with the momentum operator!---according to:
\begin{equation} \label{v}
    \bp \equiv \frac{\bJ}{|\psi|^2} = \frac{\hbar}{i} \left(\frac{\nabla \psi}{\psi} - \frac{i \bA}{\hbar}\right) = \nabla S - \bA
\end{equation}
where we have introduced an action-like quantity $S$ defined according to
\begin{equation} \label{psiS}
    \psi = e^{iS/\hbar},
\end{equation}
thus implying
\begin{equation} \label{gradS}
\nabla S =  \frac{\hbar}{i} \frac{\nabla \psi}{\psi} = \bp + \bA.
\end{equation}
The real and imaginary part of the quantity $S$ encode the
wavefunction's phase and the density profile, respectively:
\begin{align}
\Re S/\hbar &=  \arg \psi  \label{ReS} \\
\Im S/\hbar &= -  \ln |\psi|. \label{ImS}
\end{align}

The real and imaginary part, respectively, of the complex current $\bJ$ are readily evaluated:
\begin{align} 
    \bJ_1 &= \Re \left(\frac{\hbar}{i} \psi^* \nabla_1 \psi - \bA \right) |\psi|^2 \nonumber 
    \\ &= - \frac{\hbar}{2} \nabla_2 |\psi|^2 - \Re \bA  |\psi|^2, \label{J1}
\end{align}
\begin{equation} \label{J2}
    \bJ_2 = - \frac{\hbar}{2} \nabla_1 |\psi|^2 - \Im \bA |\psi|^2.
\end{equation}
Both are generally non-vanishing. 

We reiterate that Eq.~(\ref{contC}) operates expressly on the density
and, thus, constitutes a true continuity equation. This continuity
equation contains sources and sinks in the presence of external
potential. To see this more directly, one may rewrite the equation
explicitly in terms of the current's real and imaginary
components. First, substitute $\nabla = \nabla_1 - \nabla^*$ in
Eq.~(\ref{contC}) to obtain, after some algebra,
\begin{equation} \label{contC1}
    \frac{\partial \rho}{\partial t} + \frac{1}{2} \nabla_1 \bJ = -\frac{i}{\hbar}\left( \frac{|\bp|^2}{2m} + V \right) \rho - \frac{\Im \bA}{\hbar} \, \bJ.
\end{equation}
We note that $|\bp|^2  = \bp^* \bp$. Writing out the real and imaginary parts of Eq.~(\ref{contC1})  yields, respectively:
\begin{equation} \label{contj1}
    \frac{\partial \rho }{\partial t_1} + \nabla_1 \bJ_1 = \frac{2}{\hbar}\left\{ \Im V - (\Im \bA) \frac{\Re \bp}{m} \right\} \: \rho,
\end{equation}
and
\begin{equation} \label{contj2}
    \frac{\partial \rho }{\partial t_2} - \nabla_1 \bJ_2 = \frac{2}{\hbar}\left\{ \frac{|\bp|^2}{2m} + \Re V + (\Im \bA) \frac{\Im \bp}{m} \right\} \: \rho.
\end{equation}

Like Eq.~(\ref{contC}), Eqs.~(\ref{contj1}) and (\ref{contj2}) are
valid in the complex plane. The negative sign in front of the gradient
in Eq.~(\ref{contj2}) resulted from the sign choice in Eqs.~(\ref{t})
and (\ref{t*}), which was made for consistency with the spatial
coordinates. If desired, the direction of the imaginary time can be
reversed. The appearance of an energy like quantity in the curly
brackets in Eq.~(\ref{contj2}) is not surprising, if one notices that
for a stationary wavefunction at energy $E$, $\psi(\br, t) = e^{-i
  Et/\hbar} \phi(\br)$, $\partial |\psi|^2/\partial t_2 = 2E
|\psi|^2/\hbar$.

The r.h.s. of Eq.~(\ref{contj1}) vanishes on the real axis, for a
hermitian Hamiltonian $V(\br_2=0) = V(\br_2=0)^*$, $\bA(\br_2=0) =
\bA(\br_2=0)^*$, thereby yielding the conventional current
conservation:~\cite{LLquantum, Sakurai1993Modern}
\begin{equation}
    \frac{\partial \rho}{\partial t_1} + \nabla_1 \bJ_1(\br) =
    0, \hspace{5mm} \Im V = \Im A = 0, \Im \br = 0.
\end{equation}
Yet, generally, we see that continuity equations for mass transport,
when extended into the complex plane, must contain production/decay at
a spatially distributed rate, as encapsulated in the respective
right-hand sides of Eqs.~(\ref{contj1}) and (\ref{contj2}).
Incidentally, Eq.~(\ref{contj1}) implies that for non-hermitian
Hamiltonians, production/decay takes place already on the real
axis. Non-hermitian Hamiltonians are of direct interest in ${\cal
  PT}$-symmetric quantum mechanics.~\cite{RevModPhys.96.045002}

Substituting $\bJ = \bp |\psi|^2$ into Eq.~(\ref{contC}) and using the
second equality in Eq.~(\ref{v}) yields:
\begin{equation} \label{nonst}
   - \frac{\partial S}{\partial t} = \frac{\bp^2}{2m} + V +
   \frac{\hbar}{2 i m} \nabla \bp.
\end{equation}
Aside from notational differences, Eq.~(\ref{nonst}), combined with
Eq.~(\ref{gradS}), can be regarded as a complex extension of
Messiah's~\cite{Messiah} Eqs.~(VI.17) and (VI.18). Messiah states that
the latter equations ``are strictly equivalent to the Schr\"odinger
equation.'' This is not quite correct, since those equations become
indeterminate at zeros of the wavefunction, even though the
Schr\"odinger equation remains perfectly well-behaved there. This
notion is even more relevant for Eq.~(\ref{nonst}) because the
wavefunction may exhibit zeros in the complex plane even if it is
non-vanishing along the real axis. The notion of Eq.~(\ref{nonst})
being indeterminate in isolated points on the complex plane is central
to this work.

By formally taking the gradient of Eq.~(\ref{nonst}), one can bring it
to the following form:
\begin{equation} \label{Newton}
  \left( \frac{\partial}{\partial t} + \frac{\bp}{m} \nabla \right)
  \bp = q\bE + \frac{\bp}{m} \! \times \! \frac{q \bH}{c} +
  \frac{i\hbar}{2m} \left[ \nabla^2 \bp - \nabla \! \times \! \frac{q
      \bH}{c} \right],
\end{equation}
and we have reconstituted temporarily the particle's charge $q$ and
the speed of light $c$. Here, $\bE = - \left[ \partial \bA/\partial (c
  t) + \nabla V \right]$ is the electric field, $\bH = \nabla \times
\bA$ the magnetic field. To derive Eq.~(\ref{Newton}), one needs
formulas of vector calculus $\nabla (\bp^2) = (\bp \nabla) \bp + \bp
\times (\nabla \times \bp)$, $\nabla \times (\nabla S) = 0$, and
$\nabla (\nabla \bp) = \nabla^2 \bp + \nabla \times (\nabla \times
\bp)$, and Eq.~(\ref{gradS}). In view of Eq.~(\ref{dfx1}) and $\bp$,
$S$, and $\bA$ being complex differentiable almost everywhere, by
construction, we do not have to distinguish here between $\nabla$ and
$\nabla_1$.

The reader will recognize the l.h.s. of Eq.~(\ref{Newton}) as the
(complex) material derivative of the momentum in the reference frame
moving with the (complex) velocity $\bp/m$. Thus in the limit $\hbar
\to 0$, the equation of motion represented by Eq.~(\ref{Newton}) is
equivalent to the second law of Newton written down for an effective
{\em fluid} comprised of charged, mutually non-interacting particles
moving in the complex plane with velocity $\bp/m$, subject to an
electric and magnetic field, c.f. Chapter~VI of Ref.~\cite{Messiah}
and Eq.~(17.6) of Ref.~\cite{LLclass}

The $\hbar$-containing term on the r.h.s. of Eq.~(\ref{Newton}) can be
viewed as stemming from purely quantum effects, consistent with its
vanishing in the $\hbar \to 0$ limit. The contribution $(i \hbar/2m)
\nabla^2 \bp$ effectively amounts to a diffusion of the momentum field
$\bp$ with an imaginary-valued diffusivity $i \hbar/2m$. This is, of
course, consistent with the diffusion-like behavior of the
wavefunction itself, with exactly the same effective diffusivity.


Eqs.~(\ref{Newton}) thus explicitly indicates that in the eikonal
limit~\cite{LLquantum} of Eq.~(\ref{psiS}), $\hbar \to 0$, the
quantity $\bp$ can be identified with the particle's kinematic
momentum $m \bv$~\cite{Sakurai1993Modern, feynman1963lectures}
analytically continued into the complex plane. At the same time, the
quantity
\begin{equation} \label{PpAS}
    \bP \equiv \bp + \bA  = \nabla S
\end{equation}
becomes the canonical momentum~\cite{Sakurai1993Modern} $\partial
{\cal L}/\partial \bv$, c.f. Eqs.~(16.5) and (16.10) of
Ref.~\cite{LLclass}, where $\cal L$ and $\bv$ are the classical
Lagrangian and velocity, respectively, continued into the complex
plane. (The canonical momentum is sometimes referred to as the
``generalized,''~\cite{LLclass, LLquantum} ``dynamical,'' or
``p-momentum.''~\cite{feynman1963lectures}) Consistent with this
notion, the geometric object $\nabla S$ is clearly related to the
analytically-continued quantum-mechanical momentum operator, since
\begin{equation}
    \bP = \frac{\hat \bP \psi}{\psi} = \frac{\hat \bP |\psi|^2}{|\psi|^2}, 
\end{equation}
a coordinate-dependent quantity. From here on we will refer to the
objects $\bp$ and $\bP$ as the kinematic and canonical momenta,
respectively, with the understanding that $\bp$ and $\bP$ are
quantum-mechanical analogs of the respective classical momenta that
are, nonetheless, c-numbers.

\begin{figure}
\centering
\includegraphics[width= 0.8 \columnwidth]{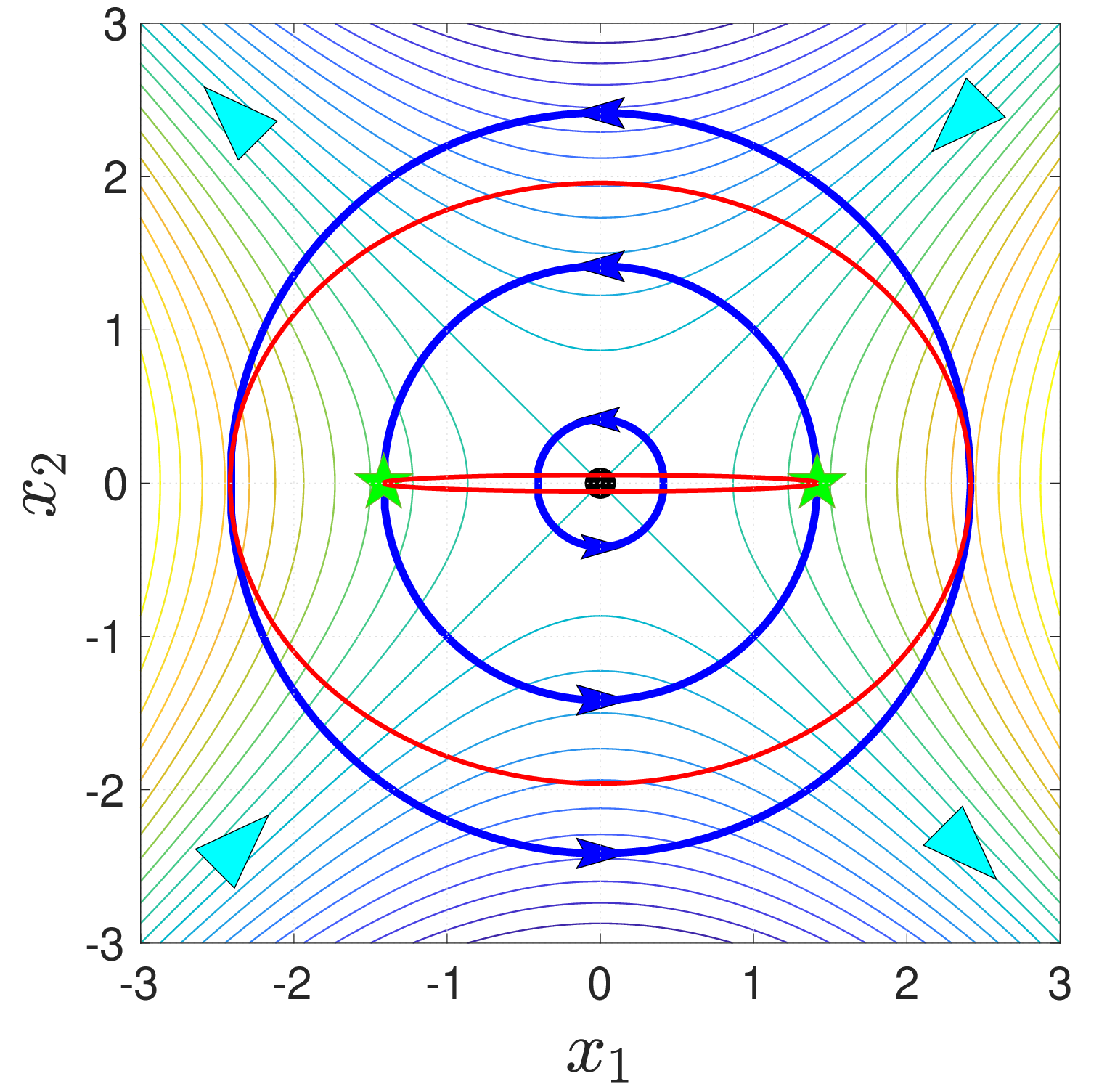}
\caption{Three blue circles with arrows illustrate trajectories $dx/p
  = d\tau$, $d \tau$ infinitesimal and real-valued, for the ground
  state of the one-dimensional Harmonic Oscillator, $V = m \omega^2
  x^2/2$, $A=0$: $p = i m \omega x$. The red ellipses show classical
  trajectories $E = p^2/2m + V$ with matching magnitudes along the
  real axis, where possible. The two stars indicate the classical
  turning points; the black dot indicates the sole critical point
  $p=0$.  The contour lines show lines of constant density $|\psi|^2$,
  warmer colors corresponding to lower densities. The cyan triangles
  indicate directions of phase flows, see Section~\ref{continuum}. $m
  = \hbar = 1/\omega = 2$.} \label{HOGS}
\end{figure}

For a stationary-state wavefunction at energy $E$, $\psi = e^{-i
  Et/\hbar} \psi(\br, E)$, Eq.~(\ref{nonst}) yields:
\begin{equation} \label{Riccati}
   E = \frac{\bp^2}{2 m} + V  + \frac{\hbar}{2 i m} \nabla \bp,
\end{equation}
which is valid in the complex plane. The complex momentum for
one-dimensional motion can be readily visualized by plotting
trajectories $dx/p = d\tau$ in the complex plane, where $\tau$ is a
real-valued parameter. We show three such trajectories for the ground
state of the harmonic oscillator in Fig.~\ref{HOGS}, using lines with
arrows. Two classical trajectories $E = p^2/2m + V$, also at energy $E
= \hbar \omega/2$, are shown in the figure as well, the lower-momentum
one being very close to the branch cut connecting the classical
turning points along the real axis. The classical trajectories can
have either orientation, while the orientation of the quantum
ones---which happens to be counterclockwise---is fixed by the choice
of the sign in front of the gradient in Eq.~(\ref{Pop}). We see the
last term in Eq.~(\ref{Riccati}) breaks the time reversal symmetry
$\bp \leftrightarrow - \bp$ of classical trajectories (continued into
the complex plane), while removing the momentum singularity at the
classical turning points, caused by the branch cut. The most
auspicious difference between the quantum momentum $p$ and the
classical momentum $p_\text{cl} = \pm [2m (E-V)]^2$ is that the former
is purely imaginary on the classically allowed portion of the real
axis---consistent with the vanishing net flux $J_1$.

Complex currents have been tacitly used by mathematicians for centuries, well predating the development of Quantum Mechanics. Indeed, Eq.~(\ref{Riccati}) can be readily presented as a real-valued equation acting on a complex-valued function:
\begin{equation} \label{RiccatiRot}
 \frac{\hbar}{2 m} \nabla_1 (i \bp)   = V - E -\frac{(i\bp)^2}{2 m}.
\end{equation}
For one dimensional motion, the above equation exemplifies the venerable Riccati equation. The connection between second order homogeneous differential equations and associated Riccati equations has been known since Euler,~\cite{10.1119/1.18535} of course, and provides a powerful tool for finding solutions to the Schr\"odinger equation for one-dimensional motion.~\cite{10.1119/1.18535, 10.1119/1.19234, RevModPhys.23.21} Thus we observe that the conventional Riccati equation, Eq.~(\ref{RiccatiRot}), expressly operates on the complex current---up to the density---not just its real part. 

Although the momentum $\bp$ itself is complex-differentiable (almost
everywhere, see below), the underlying complex current $\bJ$ is not,
since the Wirtinger divergence of its complex conjugate,
\begin{equation}
    \nabla^* \bJ = - \nabla \bJ^* = \frac{\hbar}{i} |\nabla \psi|^2,
\end{equation}
vanishes only at critical points of the wavefunction, $\nabla \psi=0$,
if any. Thus the complex current cannot be obtained by analytically
continuing some function off the real axis; instead, it should be
characterized as a complex generalization of the conventional notion
of mass current.


\section{Continuum-mechanics perspective}
\label{continuum}

According to Section~\ref{flux}, the wavefunction's property of being
complex-differentiable underlies the invariance of the Schr\"odinger
equation with respect to analytic continuation into the complex
plane. There is a conservation law associated with this local,
continuous symmetry. Indeed, complex-differentiability of a
wavefunction corresponds to the existence of an incompressible flow in
the complex plane, as appreciated early on by
Riemann.~\cite{klein1893riemann} From here on, we will be moving in
the complex plane of one spatial coordinate at a time, let it be
$x$. To simplify notation, we will drop the subscript $x$ in the
$x$-component of the kinematic momentum $\bp$: $p = p_x$, and likewise
for the $x$-components $P=P_x$ and $A=A_x$ of the canonical momentum
$\bP$ and vector potential $\bA$, respectively. Thus Eqs.~(\ref{v})
and (\ref{PpAS}) become
\begin{equation} \label{vx}
    p = \frac{\hbar}{i \psi} \frac{\partial \psi}{\partial x} - A =
    \frac{\partial S}{\partial x} - A = P - A.
\end{equation}

\begin{figure}
\centering
\includegraphics[width= 0.65 \columnwidth]{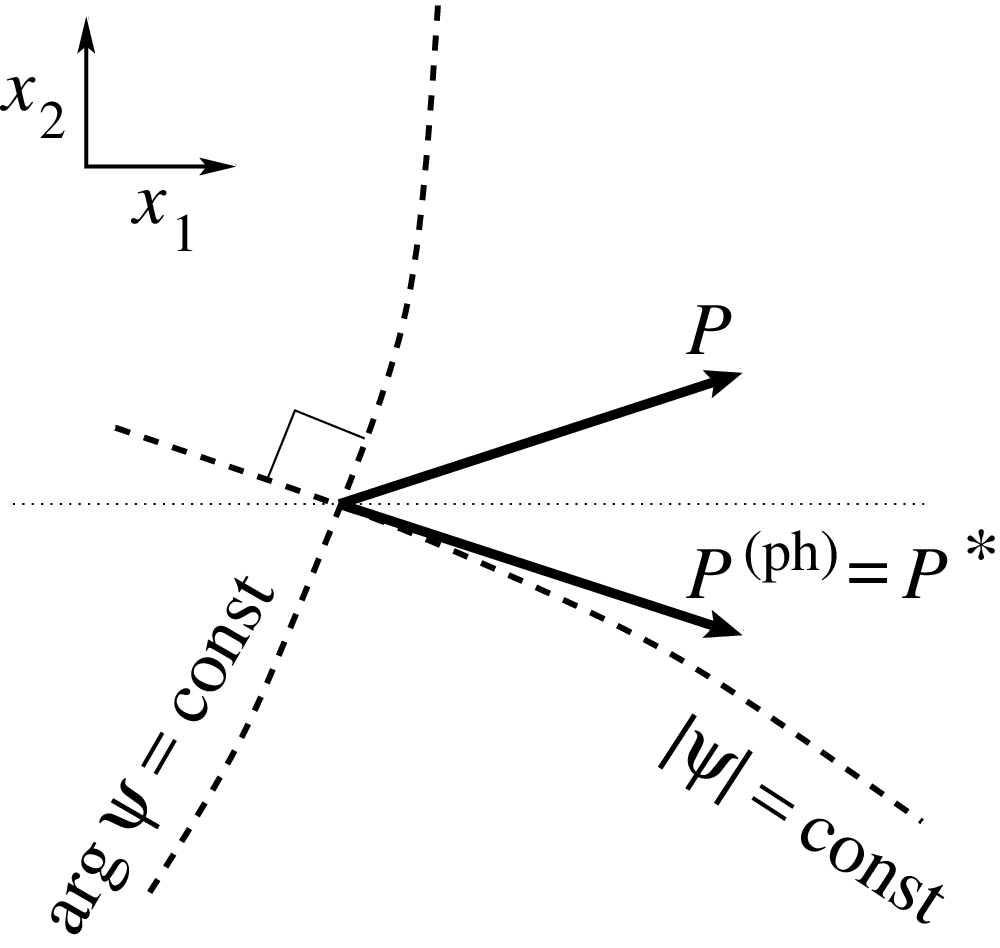}
\caption{Canonical momentum $P$ and the phase momentum $P^\ph$ in
  relation to lines of constant density and wavefunction's phase,
  respectively.} \label{pppot}
\end{figure}

Here we are specifically interested not in the wavefunction itself,
but in the closely related objects from Eq.~(\ref{vx}). We limit
ourselves to vector potentials that are holomorphic functions of the
coordinate. The complex-differentiability of $p$ implies, by
Cauchy-Riemann's conditions, that
\begin{equation}
    0 = \frac{\partial p_1}{\partial x_1} - \frac{\partial p_2}{\partial x_2} = \frac{\partial p_1}{\partial x_1} +  \frac{\partial (- p_2)}{\partial x_2}. 
\end{equation}
In other words, the object
\begin{equation}
   p^\ph \equiv p_1 - i \, p_2 = p^* 
\end{equation}
corresponds to an incompressible flow in the complex plane $(x_1,
x_2)$. Likewise, the object
\begin{equation}
   P^\ph \equiv P_1 - i \, P_2 = P^* = p^\ph + A^*,
\end{equation}
too, represents an incompressible flow in the complex plane. At the
same time, Eq.~(\ref{df*x1}) implies that
\begin{equation}
   P^* = \frac{\partial S^*}{\partial x^*} = \frac{\partial S^*}{\partial x_1} = \frac{\partial \Re S}{\partial x_1} - i \frac{\partial \Im S}{\partial x_1}. 
\end{equation}
In view of Cauchy-Riemann's conditions:
\begin{align}
    \frac{\partial \Re S}{\partial x_1} &= \frac{\partial \Im S}{\partial x_2}, \\
    \frac{\partial \Im S}{\partial x_1} &= - \frac{\partial \Re S}{\partial x_2}, 
\end{align}
one obtains
\begin{equation} \label{Sph}
   P^\ph = \frac{\partial \Re S}{\partial x_1} + i \frac{\partial \Re S}{\partial x_2}
\end{equation}
and
\begin{equation} \label{Samp}
   P^\ph = \frac{\partial \Im S}{\partial x_2} - i \frac{\partial \Im S}{\partial x_1}. 
\end{equation}
Eq.~(\ref{Sph}) means that the quantity $P^\ph$ is normal to the lines
of constant phase of the wavefunction, in the complex plane, by
Eq.~(\ref{ReS}). In other words, the momentum-like quantity $P^\ph$
mirrors the phase gradient of the wavefunction, hence the use of the
label ``ph''---referring to ``phase''---c.f. the discussion in
Ref.~\cite{Berry2013} At the same, Eq.~(\ref{Samp}) indicates that
$P^\ph$ is oriented along lines of constant density, by
Eq.~(\ref{ImS}), since $\Im S = \text{const} \Rightarrow (\partial \Im
S/\partial x_1) dx_1 + (\partial \Im S/\partial x_2) dx_2 = 0$. Thus
the lines of constant density coincide with the streamlines of the
phase momentum $P^\ph$. A graphical summary of the relation between
the canonical momentum and its phase counterpart is given in
Fig.~\ref{pppot}.

 One may consider a closed-loop integral of the canonical momentum in
 the complex plane (c.f. Chapter 10 of Ref.~\cite{LLhydro}):
\begin{eqnarray} \label{loop1}
    \oint P dx &=& \oint (P_1 + i P_2)(dx_1 + i dx_2)  \\
      &=& \oint [P_1 dx_1 + (-P_2) dx_2] \label{loop2} \\ 
      &+& i \oint (P_2 dx_1 + P_1 dx_2). \label{loop3}
\end{eqnarray}
The integral in Eq.~(\ref{loop2}) is the circulation
\begin{equation}
   \Gamma(P^\ph) = \oint [P_1^\ph dx_1 + P_2^\ph dx_2] 
\end{equation}
of the vector field $P^\ph = (P_1, -P_2)$.

The integral in Eq.~(\ref{loop3}) vanishes: 
\begin{equation}
    \oint (P_2 dx_1 + P_1 dx_2) = \oint \!\!\! \left( \!
    \frac{\partial \Im S}{\partial x_1} dx_1 + \frac{\partial \Im
      S}{\partial x_2} dx_2 \! \right) \! = \! \oint d \Im S = 0.
\end{equation}
where we used $P = (P^\ph)^*$, Eq.~(\ref{Samp}), Eq.~(\ref{ImS}), and
that the wavefunction is single-valued.

On the other hand, the integral $\oint P dx$ on the l.h.s. of
Eq.~(\ref{loop1}):
\begin{equation} \label{intdln}
  \oint P dx = \frac{\hbar}{i} \oint \frac{1}{\psi} \frac{\partial \psi}{\partial x} dx = \frac{\hbar}{i} \oint d \ln \psi,  
\end{equation}
when positively oriented, yields the number of zeros minus the number
of poles of the wavefunction contained with the integration loops,
times $2 \pi \hbar$, the zeros and poles contributing according to
their respective multiplicity. This is Cauchy's argument principle of
Complex Analysis.~\cite{ahlfors1979}

A zero of multiplicity $n$, in the wavefunction, corresponds to a pole
in the canonical momentum with residue $(+n \hbar/i)$, since
$(x^n)'/x^n = n/x$. Analogously, if the wavefunction has a pole of
multiplicity $n$ , the canonical moment acquires a pole with residue
$( - n \hbar/i)$. At the same time, Eq.~(\ref{Riccati}) dictates that
there could be two types of poles in the canonical momentum. The pole
variety of interest here is the generic kind that can arise at finite
values of the potential $V$: $|V(x)|_{|x| < \infty} <
\infty$. Substituting $P = C/(x-x_0)$, where $C$ is a constant, shows
that $C = (+1) \hbar/i$ independent of the potential $V$. (The
locations of the generic poles do depend on $V$.) These generic poles,
if any, therefore correspond to simple zeros of the wavefunction. To
give a simple example, the poles in the canonical momentum for a free
particle, $V=0$, are generic by construction. There are two linearly
independent solutions of the Schr\"odinger equation at $V=\bA=0$, a
pure cosine and sine wave respectively, that each have infinitely many
zeros on the real axis, and nowhere else.

The momentum poles of the other variety each have a residue that does
depend on the potential but the potential itself must have a pole of
the second order at the respective location, as can be checked
directly using Eq.~(\ref{Riccati}). Substituting $V = V_0/(x-x_0)^2$
into Eq.~(\ref{Riccati})---or Eq.~(\ref{SchC})---yields that the
residue of the corresponding pole in the canonical momentum is
$(\hbar/i)[1 \pm (1 + 8 m V_0/\hbar^2)^{1/2}]/2$, but would have to
also match $n \hbar/i$ in value, $n$ being a positive or negative
integer, as already mentioned. This type of momentum pole should be
regarded as accidental because it can be removed by an infinitesimal
change in the potential or mass. We will not concern ourselves with
such accidental poles in what follows.

\begin{figure}
\centering
\includegraphics[width= 0.9 \columnwidth]{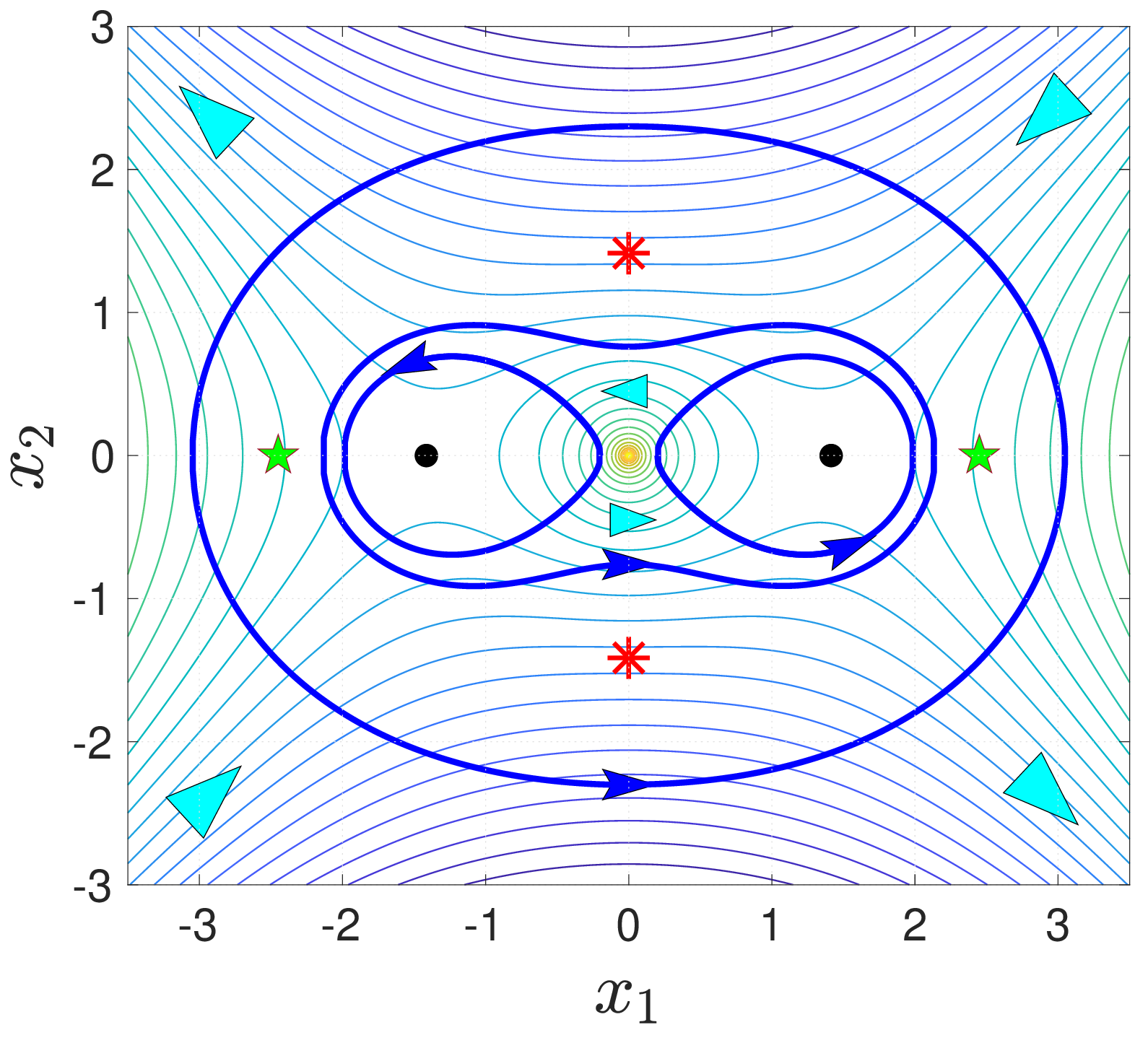}
\caption{The first excited state of the Harmonic Oscillator, $V = m
  \omega^2 x^2/2$, $A=0$. Blue loops with arrows illustrate
  trajectories $dx/p = d\tau$, $d \tau$ infinitesimal and
  real-valued. The two stars indicate the classical turning points,
  dots the locations of the critical points $\psi'=0$, asterisks the
  locations where $p_x'=0$. The contour lines show lines of constant
  density $|\psi|^2$ and, at the same time, streamlines of the phase
  momentum $P^\ph$.  The cyan triangles indicate direction of phase
  flows. $m = \hbar = 1/\omega = 2$.} \label{HO1exc}
\end{figure}

Consequently, we obtain that the circulation of the phase momentum can
be obtained by integrating, counterclockwise, the {\em canonical}
momentum along a closed loop in the complex plane. The result of the
integration is simply the number $N_0$ of the wavefunction's zeros
contained within the integration loop, times $2 \pi \hbar$:
\begin{equation} \label{qc0}
   \Gamma(P^\ph) = \oint P dx = 2 \pi \hbar \, N_0,  
\end{equation}
We note that the classical analog of the integral $\oint \bP_1 d
\br_1$ corresponds to an adiabatic invariant, Chapter 21 of
Ref.~\cite{LLclass} This is consonant with the (complex) circulation
$\oint P d x$ changing, if at all, in discrete increments. A simple
example of momentum flows accompanying a zero in the wavefunction is
afforded by the first excited state of the one-dimensional Harmonic
Oscillator, Fig.~\ref{HO1exc}.  

The notion expressed by Eq.~(\ref{qc0}) constitutes an exact
quantization condition and is closely related to the venerable
Bohr-Sommerfeld quantization condition, which is approximate: Consider
a bound state in a single potential minimum and draw a closed contour
that fully encircles the branch cut connecting the two turning points
along the real-valued classical trajectory. Let us formally substitute
into Eq.~(\ref{vx}) the WKB wavefunction $\psi_\text{WKB}(x) \sim
p_\text{cl}^{-1/2} \exp \left[(i/\hbar) \int^x p_\text{cl}(\tilde x) d
  \tilde x \right]$, where $p_\text{cl} = [2m (E - V)]^{1/2}$ is the
classical momentum, assume $A=0$, and use Eq.~(\ref{intdln}). Since
$p_\text{cl} \propto (x - x_\text{tp})^{1/2}$ near a turning point
$x_\text{tp}$, the integral of $\ln p_\text{cl}^{-1/2}$,
counterclockwise, gives $(-\pi/2) \hbar + (-\pi/2) \hbar = - \pi
\hbar$. This, then, yields the familiar expression
\begin{equation} \label{BSqcont}
  \frac{1}{2 \pi \hbar} \oint p_\text{cl} \, dx =  N_0 + \frac{1}{2}.
\end{equation}
The original derivation of Eq.~(\ref{BSqcont}) relies on the the
wave-function being single-valued;~\cite{LLquantum} consequently the
domain of the wavefunction must fit an integer or half-integer number
of wavelengths, depending on circumstances. Eq.~(\ref{qc0}), on the
other hand, provides a perspective that is more expressly topological:
The discreteness of elementary excitations stems from the number of
vortexes being inherently integer. This enables one to formulate a
quantization condition that is analogous to the Bohr-Sommerfeld
condition but, at the same time, is exact.

In many cases of practical interest, the wavefunction does exhibit
zeros on or off the real axis, or both, thus leading to a
non-vanishing circulation $\Gamma$. In other words, the phase momentum
must exhibit vortexes, in the complex plane, whose axes are each
centered at a wavefunction's zero.  This type of vortex corresponds to
a fixed point of the center type for the phase momentum, because its
streamlines coincide with the (circular) lines of constant altitude of
$|\psi| \simeq |x-x_0|$, for small values of the latter.  The
tangential velocity, $\propto 1/|\psi|$, scales inversely
proportionally with the distance to the axis of rotation and diverges
at the axis proper. The latter scaling corresponds to what they call
``irrotational vortexes'' in continuum mechanics. Indeed, the
vorticity~\cite{LLhydro, batchelor1967introduction} of the phase
momentum vanishes identically except at the very axis of rotation,
where it cannot be evaluated in the first place:
\begin{equation} \label{Omega0}
    \Omega_0^\ph = \frac{1}{m} \left(\frac{\partial P_2^\ph}{\partial
      x_1} - \frac{\partial P_1^\ph}{\partial x_2} \right) \propto
    \frac{\partial^2 \Im S}{\partial x_1 \partial x_2} -
    \frac{\partial^2 \Im S}{\partial x_1 \partial x_2} =0.
\end{equation}
and we have introduced the factor $1/m$ so that the vorticity have the
units of angular velocity, similarly to continuum mechanics. The rule
from Fig.~\ref{pppot}, then, can be used to see that each momentum
pole also represents a fixed point of the saddle type for the
streamlines of the kinematic momentum $p$.

In addition, the kinematic momentum develops rigid body-like currents
around its ``stagnation'' points $x_c$ where the momentum itself
vanishes: $p(x_c) = 0$. (The label ``$c$'' refers to ``critical,''
motivated by the notion that at $A=0$, the wavefunction's derivative
vanishes at $x_c$, by Eq.~(\ref{v}), unless the wavefunction vanishes
there, too.)  The formulas in the remainder of this Section will be
written down for one dimensional motion, but can be adjusted for
higher dimensions with the understanding that we are moving in the
complex plane for only one of the spatial coordinates at a time.

Near a critical point $x_c$, the momentum is approximately
proportional to the distance from the axis of rotation:
\begin{equation} \label{Pc}
    p(x) = i \frac{2m}{\hbar} [E - V(x_c)] (x-x_c) + o(x-x_c),
\end{equation}
by Eq.~(\ref{Riccati}), and we are limiting ourselves to stationary
solutions of the Schr\"odinger equation at energy $E$. Next we note
that the object $p/m = i \omega x = i (\omega_1 + i \omega_2) x = i
\omega_1 x + (- \omega_2) x$ is a vector---in the complex
plane---whose component oriented at $+90^\circ$ to the vector $(x_1,
x_2)$ is $\omega_1 x$, while the parallel component is $ - \omega_2
x$. In other words, $\omega_1$ is an angular velocity, while
$(-\omega_2 |x|)$ is a radial velocity. One may thus introduce the
following (complex) angular velocity corresponding to the kinematic
velocity $p/m$ near a critical point:
\begin{equation} \label{omgc}
    \omega_c = \frac{1}{m} \, \frac{p(x)-p(x_c)}{i(x-x_c)} =
    \frac{2}{\hbar}[E - V(x_c)],
\end{equation}
and $p(x_c)=0$ by construction. The vorticity corresponding to the field (\ref{Pc}) is readily evaluated:
\begin{equation} \label{reom}
    \Omega_c = \frac{1}{m} \left( \frac{\partial p_2}{\partial x_1} -
    \frac{\partial p_1}{\partial x_2} \right) = 2 \, \Re \omega_c.
\end{equation}
to yield twice the angular velocity, as expected for the vorticity of
a rigid, rotating body.~\cite{batchelor1967introduction} The quantity
$\omega_c$ generally has a non-vanishing imaginary component, too, in
which case the motion is not purely rotational but also has a radial
component:
\begin{equation} \label{radial}
   \frac{1}{m} \left( \frac{\partial p_1}{\partial x_1} + \frac{\partial p_2}{\partial x_2} \right) = - 2 \, \Im \omega_c.
\end{equation}
If so, the streamlines of the kinematic momentum become
spiral-like. In other words, the critical points are generally fixed
points of the spiral type, stable or unstable, and only become of the
center type if the potential is strictly real-valued at the axis of
the vortex, by Eqs.~(\ref{omgc}) and (\ref{radial}). Note also that
the l.h.s. is of Eq.~(\ref{radial}) is the divergence of the complex
velocity field corresponding to the kinematic momentum.  Only when the
potential $V$ is real-valued does this divergence vanish, in which
case the respective flow becomes incompressible. 

Now, in contrast with the wavefunction's zeros, there seems to be no
direct way to use closed-loop integration to directly count the
critical points $x_c$. Still, one may write down a useful weighted
sum. Consider the following closed-loop integral:
\begin{equation} \label{omegaSR}
   m \oint \frac{dx}{p} = \sum_c \frac{2 \pi}{\omega_c} = \sum_c
   \frac{\pi \hbar}{E - V(x_c)},
\end{equation}
where we used Cauchy's residue theorem and Eq.~(\ref{omgc}). The
summation is over the critical points $p(x_c)=0$. As a simple example,
consider the harmonic oscillator $V(x) = m \omega^2 x^2/2$, $\omega >
0$. Choosing the integration contour at infinity, whereby $p \to i m
\omega x$ by Eq.~(\ref{Riccati}), gives $\oint dx/(p/m) = 2\pi/\omega$
irrespective of $E$. One corollary of the sum rule in
Eq.~(\ref{omegaSR}) is that, generically, the angular frequencies
$\omega_c$ for the Harmonic Oscillator must increase linearly with the
number of the wavefunction's critical points since
\begin{equation}
    \sum_c \frac{1}{\omega_c} = \frac{1}{\omega}, \hspace{5mm} \left(
    V = \frac{m \omega^2 x^2}{2} \right),
\end{equation}
consistent with $E_n = \hbar \omega (n+1/2)$ and Eq.~(\ref{omgc}). In
any event, the sum rule in Eq.~(\ref{omegaSR}) can be used to check
consistency of approximate solutions of the Schr\"odinger equation, at
least for one-dimensional motion. It does happen to work exactly for
the particle in the box. Although the potential with infinite walls is
singular on the real axis, it can be viewed as a limiting case of a
well behaved potential. The integration contour in Eq.~(\ref{omegaSR})
should be chosen so as to avoid branch cuts in the wavefunction that
arise when the limit is taken.

One may generalize the first equality in Eq.~(\ref{omgc}) to all
points on the complex plane:
\begin{equation} \label{omgcx}
    \omega(x) = \frac{1}{i m} \, \frac{\partial p}{\partial x}.
\end{equation}
Hereby one takes a full-fledged view of the momentum $p(x)$ as a
conformal map: The quantity $\omega(x)$ still encodes rotation and
scaling, but in a local frame centered at $p(x)/m$, in the complex
plane: $p(x + dx)/m - p(x)/m = i \, \omega(x) dx$. For the conformal
map to be well-defined, it is necessary that
\begin{equation} \label{conjcond}
    \frac{\partial p}{\partial x} \ne 0,
\end{equation}
This condition can be violated, if at all, only in isolated points on
the complex plane except in the free particle case. Indeed, suppose on
the contrary that $\partial p/\partial x = 0$ along some line. By
Eq.~(\ref{Riccati}), $p^2/2m + V(x) = \text{const}$ along such a
line. Varying the latter equality with respect to $x$ yields $\partial
p/\partial x = - m V'/p$ which, however, can vanish only in isolated
points unless $V = \text{const}$ and we assume $V$ is reasonably
well-behaved. As a side dividend, we just showed that a quantum and a
classical streamline, respectively, that correspond to the same value
of energy can intersect only in isolated points. We observe that the
last term in Eq.~(\ref{Riccati}) enables the $p = p(x)$ conformal
mapping, except in isolated points on the complex plane. (This is in
addition to what we discussed immediately following
Eq.~(\ref{Riccati}).) For instance, the first excited state of the
Harmonic Oscillator violates the necessary condition (\ref{conjcond})
in two points, indicated by the red asterisks in Fig.~\ref{HO1exc}.

Now, Eq.~(\ref{Riccati}) and (\ref{omgcx}), together, imply a
suggestive expression
\begin{equation} \label{pconf}
    E = \frac{p^2}{2m} + V(x) + \frac{\hbar \omega(x)}{2}.
\end{equation}
Eq.~(\ref{pconf}) presents the total kinetic energy as a sum of
contributions due to conformal translation ($p^2/2m$) and
rotation/scaling ($\hbar \omega(x)/2$),
respectively. Eq.~(\ref{pconf}) can be generalized to multiple spatial
dimensions whereby $x \to \br$, $p \to \bp$, and $\omega(x) \to
\omega(\br) = \nabla \bp/im$.

The rotational nature of the $\hbar \omega(x)/2$ term becomes
particularly overt at large $x$. (We stipulate by construction that
bounding or scattering potentials are centered at finite separation
from the origin.) Under these circumstances, the last term in
Eq.~(\ref{Riccati}), $\hbar p'/2 i m$, can be made arbitrarily smaller
than the totality of the remaining terms. This is formally equivalent
to taking the $\hbar \to 0$ limit. Within the leading order in
$\hbar$, the quantity $p'$ inside the $\hbar p'/2 i m$ term can be
replaced by its value $p' \approx -m V'/p$ at $\hbar=0$: 
\begin{equation} \label{RiccatiSC}
  E = \frac{p^2}{2m} + V + i \hbar  \frac{m V'}{2 p} + o(\hbar).
\end{equation}
Indeed, at $\hbar = 0$ (and fixed $y$ and $z$, if in three
dimensions), $p^2/2m+V = \text{const} \Rightarrow (p/m)dp + V'dx =
0$. Thus at large $x$ one obtains asymptotically, by
Eqs.~(\ref{pconf}) and (\ref{RiccatiSC}):
\begin{equation} \label{omVp}
    \omega(x) = i \frac{V'}{p}.
\end{equation}
But this equality happens to formally express the
force balance between the centrifugal and centripetal forces,
respectively, acting on a test particle of mass $m_0$ orbiting in the
complex plane with velocity $i \omega(x) x = p/m_0$:
\begin{equation} \label{balance}
    m_0 \omega^2(x) \: x - V' = \frac{p}{i} \omega(x) - V' = 0.
\end{equation}

For polynomial potentials of leading order $n$, Eqs.~(\ref{pconf}) and
(\ref{omVp}) yield $\omega(x) \propto x^{n/2-1}$ for large $x$. The
quadratic potential $n=2$---stable or inverted---is special in that
its rotational frequency $\omega(x)$ tends to a steady value at
infinity, implying the kinematic momentum displays a rigid body-like
flow pattern at infinity. For any bound state of the harmonic
oscillator $V = m \omega^2 x^2/2$, $\omega(x)_{x \to \infty} \to
\omega >0$. The ground state of the harmonic oscillator is
particularly simple since its momentum $p = i m \omega x$, yielding
$p^2/2m + V(x) = 0$ and $\omega(x) = \omega$ throughout. In other
words, the zero-point energy is due to a rigid rotation of the
momentum field in the complex plane at large $x$;
c.f. Figs.~\ref{HOGS} and \ref{HO1exc}.

Lastly we note that the conformal-rotational energy $\hbar \omega/2$
of the harmonic oscillator matches, in value, the rotational energy of
a classical particle whose angular momentum and angular velocity are
given by $\hbar$ and $\omega$, respectively. The value of the orbital
momentum associated with the zero-point vibration of the oscillator is
apparently quantized. This can be understood using symmetry
considerations, see Appendix~\ref{rays}.

\section{Complex momentum flows}
\label{flows}

In this Section, we use available solutions of the Schr\"odinger
equation for one-dimensional motion to illustrate how momentum flows
come about in the complex plane, in the context of physical phenomena
of interest. We will highlight key features of such flows; a more
in-depth yet non-technical discussion of these features can be found
in Appendix~\ref{rays}. The computational details are provided in
Appendix~\ref{details}.

\begin{figure}
\centering
\includegraphics[width= 0.85 \columnwidth]{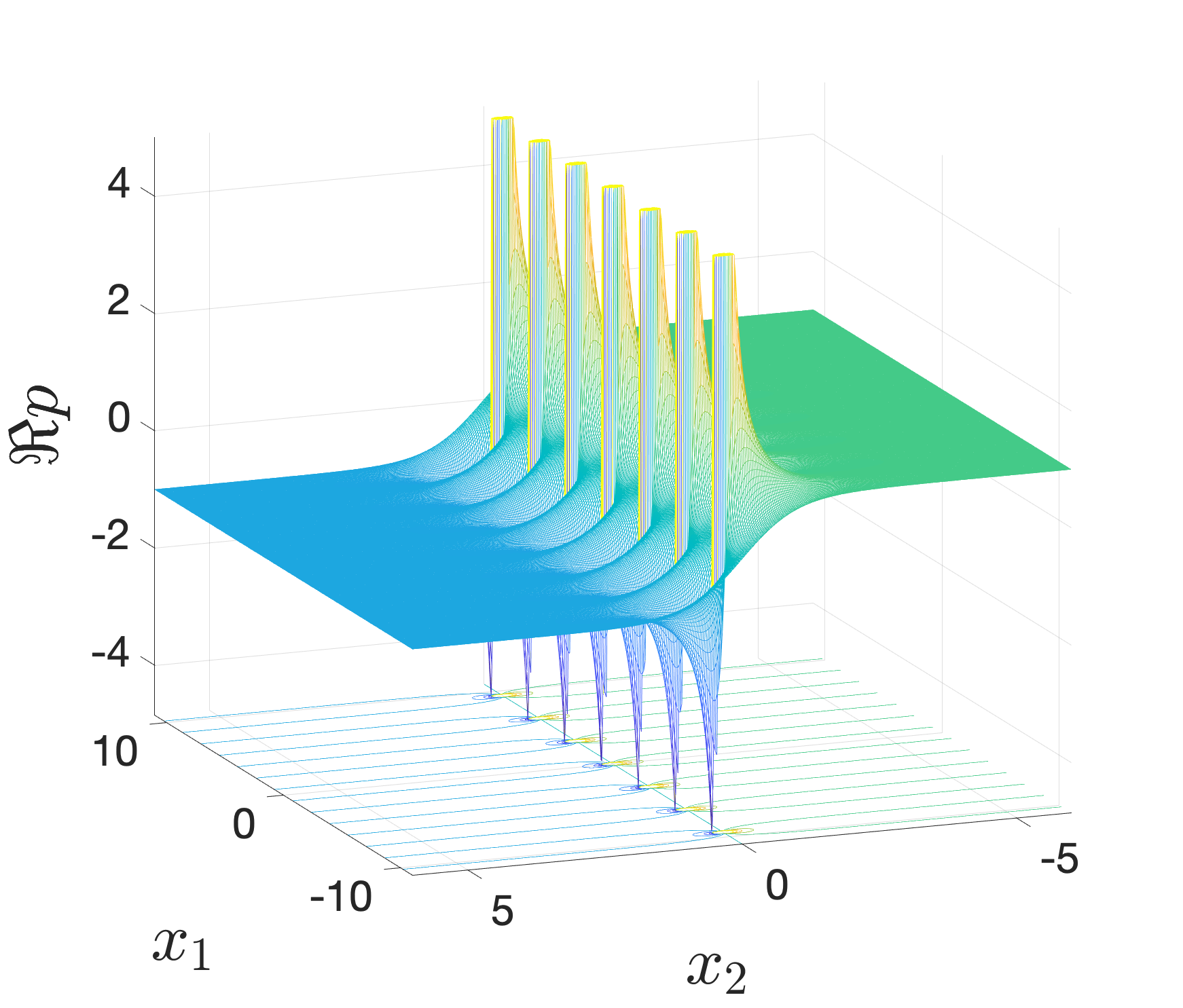}
\caption{The real part of the complex momentum $p = (\hbar/i)
  \text{cotan}(x)$ corresponding to the free particle  $\psi =
  \sin(kx)$. At those $x$, where $|p| > 6$, we set $p =6$ for the sake
  of presentation. $m = \hbar = 2$, $E = 1$.  } \label{FPFig}
\end{figure}

We have already considered two simple cases of a standing wave, with
no nodes in Fig.~\ref{HOGS}, and with just one node in
Fig.~\ref{HO1exc}. The opposite extreme of a fully extended standing
wave is exemplified by a real-valued solution for the free particle,
make it $\psi \propto \sin(k x)$ for concreteness, $k$ real-valued and
positive. We show the real part of the corresponding momentum $p =
(\hbar/i) \text{cotan}(x)$ in Fig.~~\ref{FPFig}. We observe a string
of poles spaced at $\pi/k$ along the real axis, each corresponding to
a zero of $\psi$. As one moves away from the real axis, the momentum
becomes well behaved while approaching its classical value $\pm
\sqrt{2m E}$. The transition from the singular behavior near the poles
to the smooth, classical-like momentum flow takes place exponentially
fast, the pertinent length scale given by the spacing between adjacent
poles, up to a constant of order one. For instance at $x_1 = 0$,
$(1/i)\text{cotan}(k x) = -\text{cotanh}(k x_2) \simeq - (1 - e^{-2 k
  |x_2|}) \text{sign} (x_2)$, for $|x_2| \gtrsim 1/k$.

Next we consider scattering by the inverted, one-dimensional parabolic
potential, $V = - m \omega^2 x^2/2$, $A=0$, and we adopt $\hbar = m =
1/\omega =2$ as in Ref.~\cite{BARTON1986322} Specifically we are
interested in stationary solutions of the Schr\"odinger equation such
that the incident and reflected signal are on the l.h.s. of the
barrier, by construction, while the transmitted signal is on the
r.h.s. of the barrier and is in the form of a single wave propagating
to the right.  There are two rather distinct cases to consider,
corresponding to energies below and above the top of the barrier,
referred to as ``overdense'' and ``underdense,''
respectively.~\cite{BerryMount1972} The parabolic potential turns out
to exhibit a special symmetry whereby the overdense and underdense
cases are mutually complimentary, when considered in the complex
plane. Indeed, Eqs.~(\ref{dx}) and (\ref{dx*}) imply that $\partial^2
\psi/\partial x^2 = \partial^2 \psi/\partial x_1^2 = - \partial^2
\psi/\partial x_2^2$.  Consequently, the energy for motion along the
real axis is the negative of the energy for motion along the imaginary
axis:
\begin{equation}
    \left( - \frac{\partial^2}{\partial x_1^2} - \frac{x_1^2}{4} \right) \psi = E \psi \Leftrightarrow \left( - \frac{\partial^2}{\partial x_2^2} - \frac{x_2^2}{4} \right) \psi = - E \psi.
\end{equation}
Furthermore, according to Appendix~\ref{rays}, if we stipulate that
the solution along the positive real half-axis be an outgoing wave,
the solution along the positive imaginary half-axis must be an
outgoing wave as well.

\begin{figure}
\centering
\includegraphics[width= 0.85 \columnwidth]{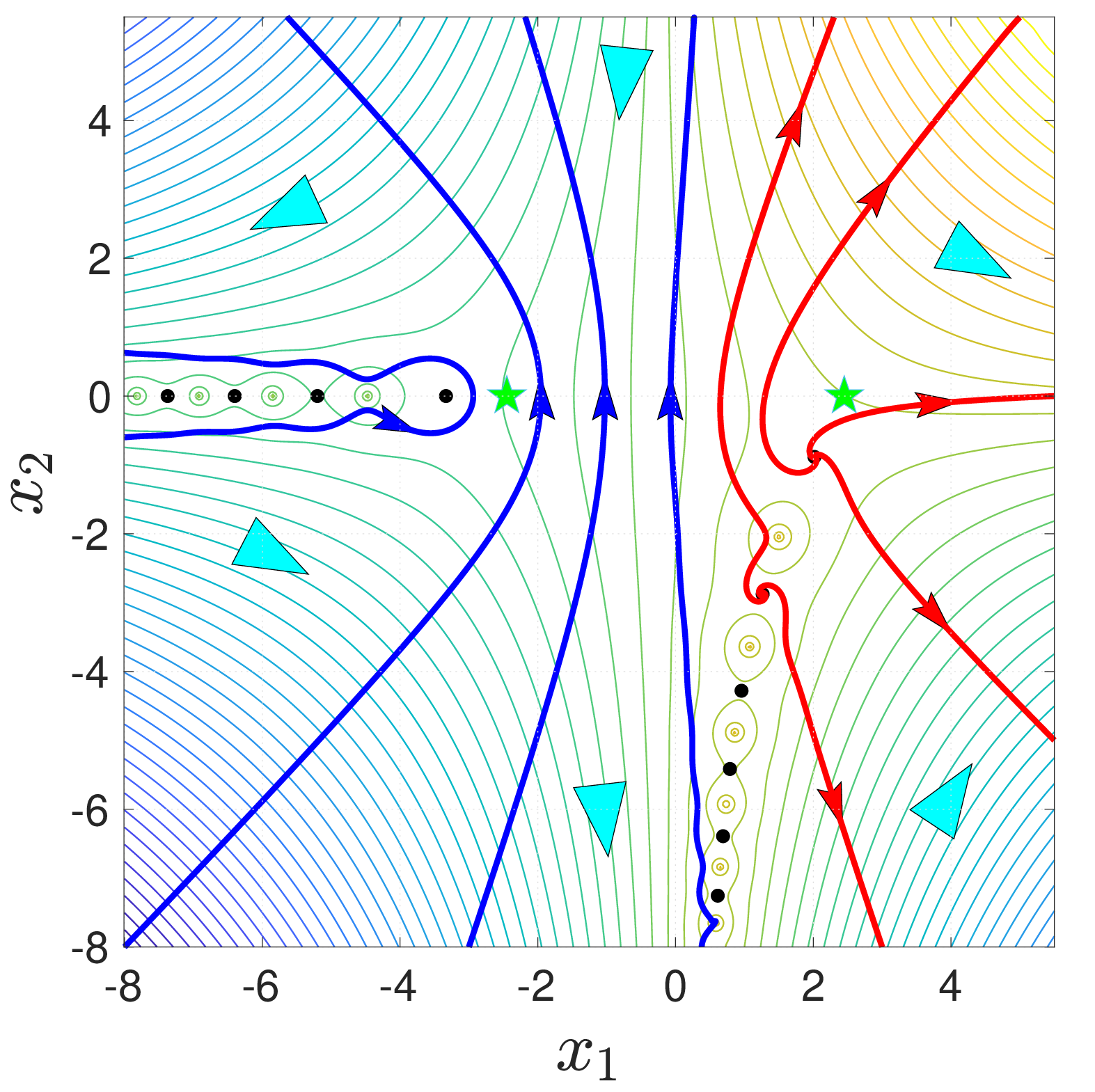}
\caption{Overdense scattering by the inverted Harmonic Oscillator, $V
  = -m \omega^2 x^2/2$, $E = - 1.5 \hbar \omega$. Blue and red thick
  lines with arrows illustrate trajectories $dx/p_x = d\tau$, $d \tau$
  infinitesimal and real-valued. The two stars indicate the classical
  turning points, dots the locations of the critical points
  $\psi'=0$. The contour lines show lines of constant density
  $|\psi|^2$ and, at the same time, streamlines of phase-momentum
  flows. The cyan triangles show direction of the phase flows. $m =
  \hbar = 1/\omega = 2$.} \label{TunnFig}
\end{figure}

We begin with the overdense case, which explicitly involves tunneling.
The key features of the wavefunction and the corresponding complex
flows are graphically summarized in Fig.~\ref{TunnFig}. The energy
value $E = - 1.5 \hbar \omega$ was chosen for graphical clarity, the
corresponding transmission coefficient~\cite{PhysRev.48.549,
  LLquantum, BerryMount1972} being approximately $T \simeq e^{-2 \pi
  |E|/\hbar \omega} \approx 8.1 \cdot 10^{-5}$.

The momentum for this solution exhibits two infinite strings of
poles. The poles are easily spotted in the figure for they must be
contained within closed streamlines of the phase momentum $P^\ph$. The
local direction of the phase flow is indicated by the pointed cyan
triangles. When occurring along closed loops, the flow of the phase
momentum is always counterclockwise, according to
Section~\ref{continuum}, c.f. Fig.~\ref{HO1exc}. Because the closed
loops are located only along strings of poles, the direction of the
phase flow must reverse across a string of poles but remains smooth
otherwise and is well approximated by (the complex conjugate) of the
classical momentum.

The string of poles running along the negative side of the real axis
reflects the interference between the incoming and reflected
signal. These poles are each shifted upwards off the real axis: On the
one hand, the adjacent portion of the real axis is in the classically
allowed region, and so the rigid-body currents around the critical
points must be counterclockwise, by Eq.~(\ref{omgc}) and
(\ref{reom}). On the other hand, the mass current along the real axis
is positive by construction.  (The critical points are indicated with
black dots.) By Eq.~(\ref{J1}), the magnitude of the shift of the
poles off the real axis scales with the transmission coefficient and
is therefore small, but non-vanishing nonetheless. The horizontal
location of these poles determines the phase shift of the reflected
signal, relative to the incoming signal. The significance of the
string of poles running along the imaginary axis will become clear
shortly.

The streamlines of the kinematic momentum are illustrated using select
trajectories $dx/p = d\tau$, $d\tau$ infinitesimal and real-valued;
these trajectories are denoted with thick blue and red lines,
respectively, each marked with an arrow. The blue streamlines originate
from the third quadrant and pertain to the incoming signal. The red
streamlines show select trajectories arriving in the 1st and 4th
quadrant; these pertain to the transmitted signal.  We observe that
the incoming signal is largely diverted toward the second quadrant,
while the streamlines for the signal on the receiving end originate
from critical points located on the r.h.s. of the barrier, i.e. the
side opposite of the incident signal.

\begin{figure}
\centering
\includegraphics[width= 0.85 \columnwidth]{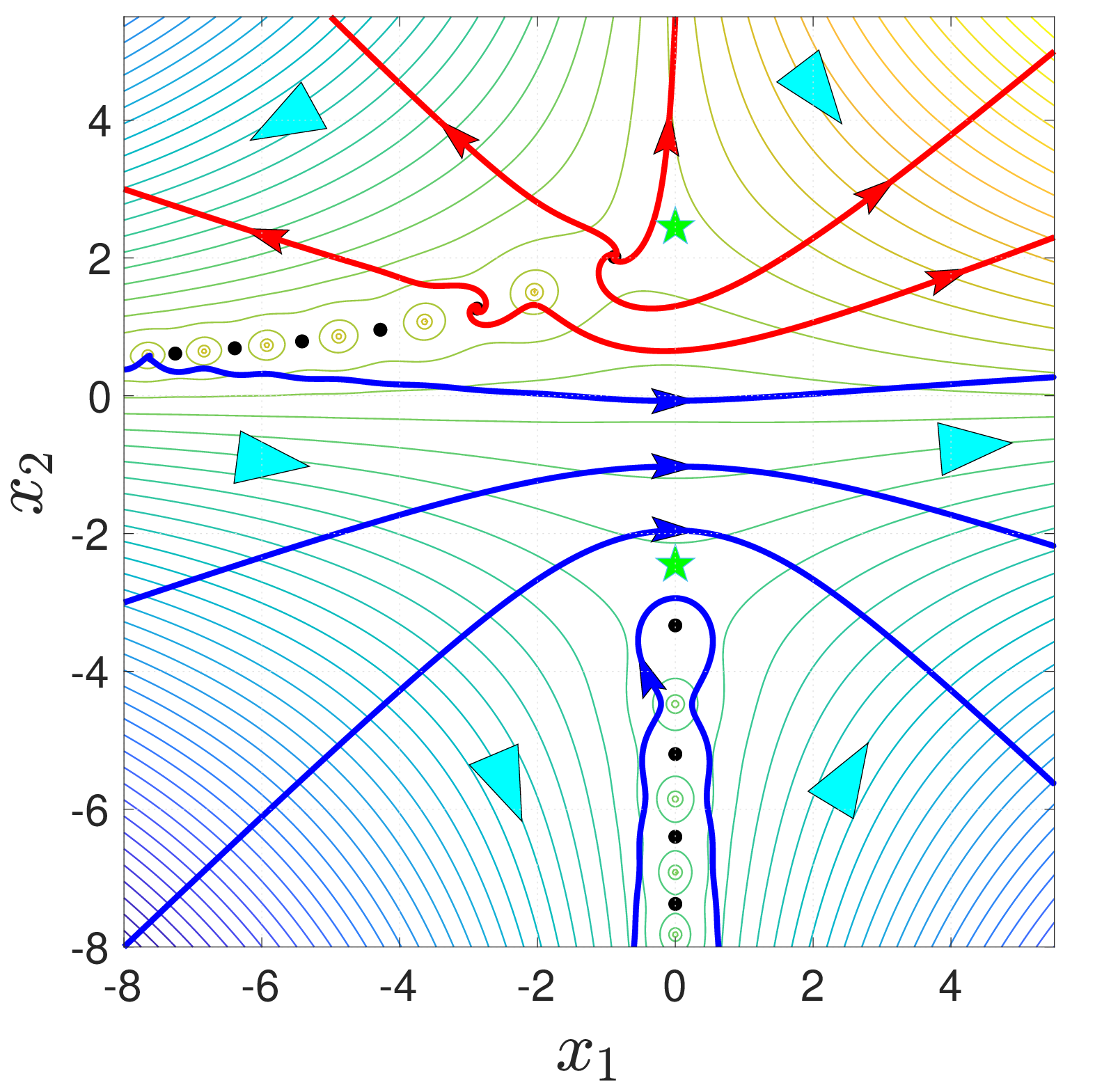}
\caption{Underdense scattering by the inverted Harmonic Oscillator, $V = -m \omega^2 x^2/2$, $E = + 1.5 \hbar \omega$. The rest of the notation is the same as in Fig.~\ref{TunnFig}.} \label{ScatFig}
\end{figure}

The direction of the rigid-body flows of the kinematic momentum,
around the critical points, depends on the points' locations. We have
already commented on the set of the critical points located in the
second quadrant. The set located in the forth quadrant corresponds to
classically forbidden motion. According to Eq.~(\ref{omgc}) and
(\ref{reom}), then, the orientation of the rigid-body flows for this
set must be clockwise. For both sets, Eqs.~(\ref{omgc}) and
(\ref{radial}) dictate that the radial component of the momentum flow
be positive, since $x_1 x_2 < 0$ in the second and fourth
quadrants. Thus all of the critical points in Fig.~\ref{TunnFig} must
be fixed points of the unstable-spiral type, confirming the trend
exhibited by the red trajectories in Fig.~\ref{TunnFig}, see also
Fig.~\ref{MapCurrentsFig} below. That trajectories begin in these
critical points is consistent with the second and fourth quadrants
being particle sources, in view of Eq.~(\ref{contj1}) and $\Im (-x^2)
> 0$ in these quadrants, see also the comment following
Eq.~(\ref{radial}).

The analysis for the underdense case, illustrated in
Fig.~\ref{ScatFig}, can be performed analogously. We choose energy $E=
+ 1.5 \hbar \omega$, for the sake of argument, even though the
reflection coefficient is hereby rather small. Owing to the symmetry
discussed in the beginning of this Section, the features in
Fig.~\ref{ScatFig} are mirror images of the features from
Fig.~\ref{TunnFig} about the $x_2 = x_1$ line, except for the
directions of the phase flows. We observe that at this low amount of
reflection, not only does the mass flux along the real axis become
substantial but most of the transmitted signal on the real axis
originates from the third quadrant, as one would expect. The large
amount of flux along the real axis must be matched by a substantial
shift upwards of the string of poles running along the real axis, per
Eq.~(\ref{J1}). At the same time, the poles along the imaginary axis
are shifted down to be below the pertinent turning point. Informally
speaking, these two circumstances amount to opening a passage for the
streamlines along the real axis. In retrospect, the visible bending of
the string of poles off the imaginary axis in Fig.~\ref{TunnFig} was
to provide for a substantial particle flux along the imaginary axis.

\begin{figure}
\centering
\includegraphics[width= 0.85 \columnwidth]{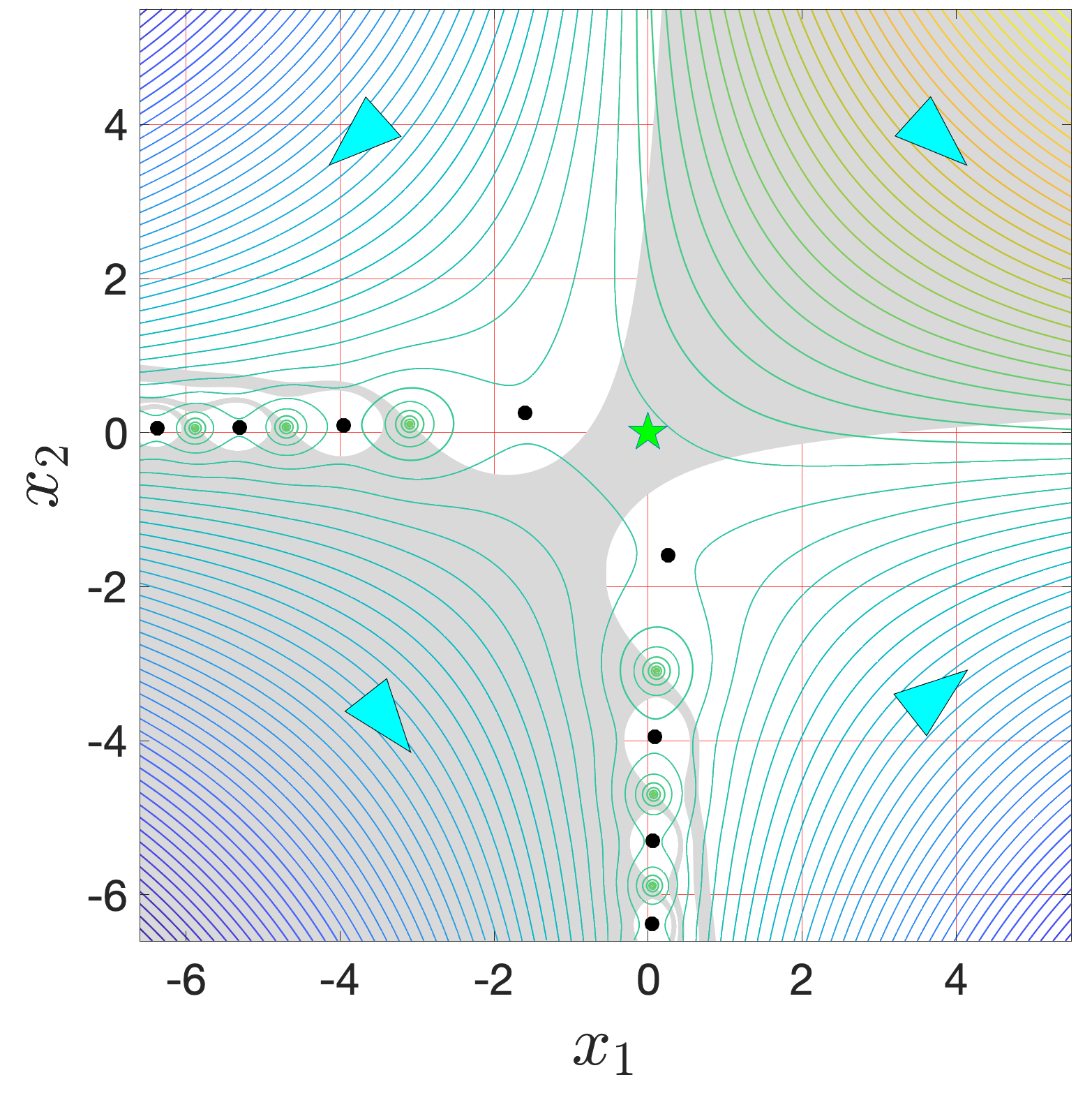}
\caption{Scattering for the marginal case $E=0$ of the inverted
  Harmonic Oscillator, $V = -m \omega^2 x^2/2$. The shaded areas
  represents the totality of the momentum streamlines originating in
  the third quadrant. The contour lines show lines of constant density
  $|\psi|^2$ and, at the same time, streamlines of the
  phase-momentum. The cyan triangles show the direction of the
  respective flows. The green star shows the location of the sole
  classical turning point. $m = \hbar = 1/\omega =
  2$.} \label{MapCurrentsFig}
\end{figure}

It may appear that a string of poles that ends on approach to a
classical turning point tends to bend so as to extrapolate to that
turning point. This is only approximately so, as can be directly seen
by considering the important marginal case $E=0$ for the same
inverted-parabola scattering potential. At $E=0$, the problem exhibits
an additional symmetry, whereby the classical turning points become
merged into one, located at the origin. The lines of constant density
then exhibit a mirror symmetry under the reflection about the $x_2 =
x_1$ line, see Fig.~\ref{MapCurrentsFig}. Consequently, the
streamlines of the phase momentum that originate at the actual signal
source never leave the third quadrant, as is the case for $E<0$.

The value $E=0$ presents additional convenience in that it allows one
to readily see that the critical points along both strings of poles
are sources. The shaded area in Fig.~\ref{MapCurrentsFig} is the
conglomerate of all trajectories that originate from the third
quadrant. The empty area corresponds, then, to trajectories that
originate at the critical points. We observe that already at $E=0$,
the streamlines for the transmitted signal originate on the outgoing
side of the barrier, not at $x_1 = - \infty$.

\begin{figure}
\centering
\includegraphics[width= 0.9 \columnwidth]{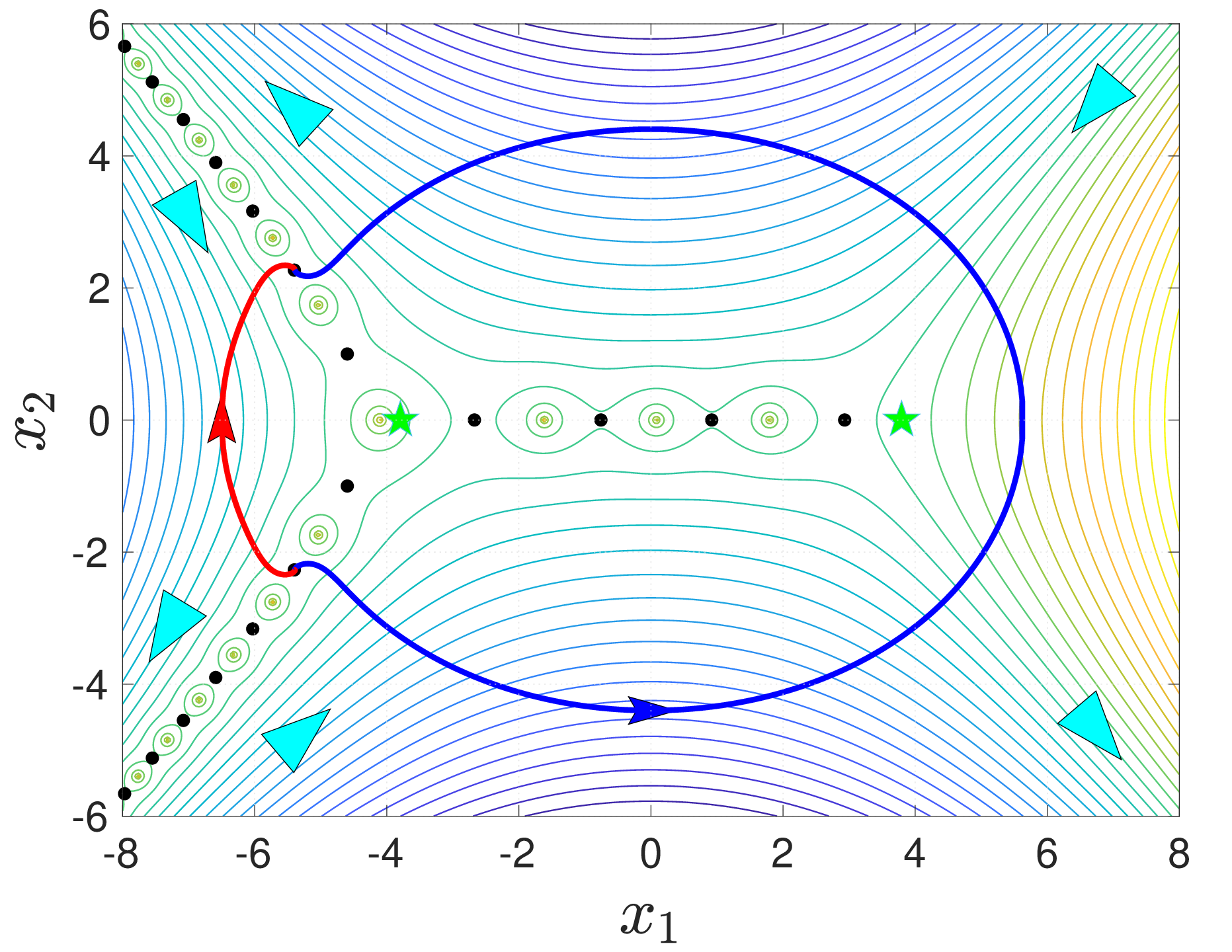}
\caption{An (unnormalized) solution for the Harmonic Oscillator, $V =
  m \omega^2 x^2/2$, $A=0$, at $E = 3.6 \hbar \omega$. The blue and
  red streamlines with arrows each exemplify a trajectory $dx/p =
  d\tau$, infinitesimal and real-valued. The two stars indicate the
  classical turning points, dots the locations of the critical points
  $\psi'=0$. The contour lines show lines of constant density
  $|\psi|^2$ and, at the same time, streamlines of the
  phase-momentum. The cyan triangles indicate direction of phase
  flows. $m = \hbar = 1/\omega = 2$.} \label{HOunNorm}
\end{figure}

The patterns of momentum flows in the complex plane are directly
connected to the normalizability of the wavefunction. We illustrate
this notion by showing the solution of the Schr\"odinger equation for
the Harmonic Oscillator at an energy other than $\hbar \omega
(n+1/2)$, $n = 0, 1, 2 \ldots$. Under these circumstances, it is still
possible to normalize the wavefunction along either the positive or
negative direction of the real axis, but not both at the same time. As
a result, two strings of poles emerge to flank the real axis on the
unnormalized side, see Fig.~\ref{HOunNorm}; concurrently, the momentum
flows on the unnormalized side reverse to become directed
clockwise. These emerged strings of poles happen to run along
anti-Stokes lines, see Appendix~\ref{rays}. One may, then, tailor the
quantization condition (\ref{qc0}) to help detect whether a
wavefunction is normalized, by stipulating that the vertical extent of
the area covered by the integration loop is large enough to contain
the anti-Stokes lines adjacent to the pertinent side of the real
axis. The extent should be, however, not to so large as to cover other
anti-Stokes lines or the wavefunction's singularities off the real
axis, if any.

\section{Momentum in the complex plane as a distributed quantity}
\label{fluct}

Consider the following object:
\begin{equation}
    \zeta(\br) \equiv |\psi_A(\br)|^2,
\end{equation}
where the quantity $\psi_A$ is defined according to:
\begin{equation} \label{psiA}
    \psi_A(\br) \equiv \psi(\br) e^{-\frac{i}{\hbar} \int^\br \bA(\tilde \br) d \tilde \br}.
\end{equation}
It is a gauged wavefunction that solves Eq.~(\ref{SchC}) modified according to $(- i \hbar \nabla - \bA) \to (-i \hbar \nabla)$.

Let us introduce the following quantity:
\begin{eqnarray} \label{C1}
    C_1 &=& \frac{1}{i} \frac{\partial \ln \zeta}{\partial x} =
    \frac{1}{i \zeta} \frac{\partial \zeta}{\partial x} = \frac{1}{i
      \psi_A} \frac{\partial \psi_A}{\partial x} \\ &=& \frac{1}{i
      \psi} \frac{\partial \psi}{\partial x} - \frac{A}{\hbar} =
    \frac{p_x}{\hbar}. \nonumber
\end{eqnarray}
In assessing these equations, it is useful to think of the wavefunction $\psi_A$ as a wavepacket of sorts:
\begin{equation} \label{psiAwp}
    \psi_A(\br) \propto \int e^{i \tilde \bp \br/\hbar} \tilde \psi_A (\tilde \bp) \, d^3 \tilde \bp.
\end{equation}
For each of the spatial variables, the integration is along some
contour in the respective complex plane. Incidentally, this type of
expansion is used in Laplace's contour integration method of solving
differential equations, see Appendices \S a, b, and d of
Ref.~\cite{LLquantum}, or Ref.~\cite{TFKP} Within the construct
(\ref{psiAwp}), the quantity $C_1$ becomes a first cumulant:
\begin{equation} \label{C1A}
    C_1 = \frac{\int \tilde p_x e^{i \tilde \bp \br/\hbar} \tilde \psi_A (\tilde \bp) \, d^3 \tilde \bp}{\int e^{i \tilde \bp \br/\hbar} \tilde \psi_A (\tilde \bp) \, d^3 \tilde \bp} = \langle \tilde p_x \rangle,
\end{equation}
where we used the third equality in Eq.~(\ref{C1}) and the angular
brackets denote averaging. The tilde on the r.h.s. signifies that the
averaging pertains to the ``generating function'' $\tilde \psi_A$.
Similarly to how it is done in Thermodynamics, in which the partition
function serves as the moment-generating function, here we do not set
the arguments $\br$ of our generating function $\psi_A$ to zero after
differentiating it. Instead, we retain them as variables to enable us
to generate their conjugate counterparts. To this end, consider an
effective potential:
\begin{equation}
    F(\br) = - \hbar \ln \zeta(\br).
\end{equation}
Eq.~(\ref{C1}) implies that the component $p_x$ of the momentum can be thought of as conjugate to the variable $(i \, x)$ and, at the same time, that $p_x$ is equal to the expectation value of the argument $\tilde p_x$ of the (weighted) distribution defined by the integrand in Eq.~(\ref{psiAwp}):
\begin{equation} \label{px}
    p_x = - \frac{\partial}{\partial (i \, x)} F = \langle \tilde p_x \rangle.
\end{equation}
For the second cumulant, one readily obtains:
\begin{eqnarray} \label{C2}
    C_2 &=& - \frac{\partial^2 }{\partial x^2} \ln \zeta =  \frac{1}{i} \frac{\partial }{\partial x} C_1   = \frac{1}{i \hbar} \frac{\partial }{\partial x}  p_x \\ &=& \frac{ \langle \tilde p^2_x \rangle - \langle \tilde p_x \rangle^2}{\hbar^2}.  \label{C2b}
\end{eqnarray}
One may define the first and second moments, respectively, in a normalization-independent way: 
\begin{equation} \label{M1}
    M_1 = \frac{1}{\psi_A} \frac{\partial \psi_A}{i \partial x} = C_1,
\end{equation}
\begin{equation} \label{M2}
    M_2 = - \frac{1}{\psi_A} \frac{\partial^2 \psi_A}{\partial x^2} = \langle \tilde p^2_x \rangle.
\end{equation}
One immediately observes that $M_2 = M_1^2 + C_2$---as it should be---and, furthermore, may readily verify that the second moment also yields the kinetic energy up to a numerical factor:
\begin{equation} \label{M2kin}
    M_2 = M_1^2 + C_2 = \frac{1}{\hbar^2}\frac{\left(-i \hbar \nabla - \bA \right)^2 \psi}{\psi}.
\end{equation}

According to Eq.~(\ref{C1}) and (\ref{C2}), the increment of the potential $F$  
\begin{equation} \label{Fxdx}
    F(x+ \Delta x)/\hbar  = F(x)/\hbar  - i C_1 \Delta x + \frac{C_2}{2}\Delta x^2 + o(\Delta x^2),
\end{equation}
where $C_1$ and $C_2$ are evaluated at $x$. According to
Eq.~(\ref{Fxdx}), the quantity $\hbar C_2$ is a response function, 
c.~f. the last equality in Eq.~(\ref{C2}). Since $\zeta =
e^{-F/\hbar} \propto e^{-C_2 \Delta x^2/2}$, the mean square deviation
of $x$ is
\begin{equation} \label{deltax2}
    (\delta x)^2 = \frac{1}{C_2},
\end{equation}
where the direction of steepest descent is along the line
$-\arg(C_2)/2$ in the complex plane. We have thus determined an
effective potential that directly determines fluctuations of the
coordinate. The word ``fluctuation'' refers to the value of $x$ being
distributed, without any implication of stochasticity whatsoever.

At the momentum poles, the response function $C_2$ diverges, implying
the fluctuation of the coordinate, Eq.~(\ref{deltax2}), become
vanishing.  This is a quantitative way to express that the relative
phases of individual waves comprising a standing wave become strictly
fixed. Conversely, the response function $C_2$ vanishes at locations
where $F'' =0$, i.e., at spinodals~\cite{HLKnowledge} of the effective
potential $F(x)$. In thermodynamics, we associate spinodals with
mechanical stability limits of a thermodynamic phase. Note it is at
these spinodals that the condition (\ref{conjcond}) for the mapping
$p_x(x)$ to be conformal is violated.

Eq.~(\ref{C2b}) suggests a way to establish an effective potential
that determines the average value and fluctuations, respectively, of
the momentum.  According to the first equality in Eq.~(\ref{px}), one
may formally define a Legendre transform
\begin{equation} \label{Gpx}
    G(p_x) = F + i x p_x,
\end{equation}
where $x$ is considered a function of $p_x$ through the first equality
in Eq.~(\ref{px}).  This is possible within any region of the complex
plane $(x_1, x_2)$ such that the function $x(p_x)$ is univalent. When
the wavefunction exhibits zeros, the function $x(p_x)$ however becomes
multi-valued. Already when one pole is present---as is the case for
the first excited state of the Harmonic Oscillator,
Fig.~\ref{HO1exc}---the function $x(p_x)$ is two-valued for almost all
values of $p$. Indeed, $p = - 2 i (1/x - x/2) = 2^{3/2} [(i
  x/\sqrt{2})^{-1} + (i x/\sqrt{2})]$, which happens to be the
venerable Joukowsky transform, up to rotation and scaling. A conformal
map $p_x \leftrightarrow x$ can still be defined separately for
mutually-complementary regions on the $x$ plane. One can use, for
example, the pair of regions $|x| < \sqrt{2}$ and $|x| > \sqrt{2}$,
respectively.~\cite{TFKP, ahlfors1979}

Subject to this caveat, the quantity $G(\bp) = F + i \, \br \bp$,
where $\br = \br(\bp)$, then formally defines a generating function
with the momentum $\bp$ as its argument:
\begin{equation}
    \xi(\bp) = e^{-G(\bp)/\hbar}.
\end{equation}

Combining Eq.~(\ref{px}), (\ref{Fxdx}), and (\ref{Gpx}) yields:
\begin{equation} \label{Gp}
    G(p_x+ \Delta p_x)/\hbar - G(p_x)/\hbar = \frac{1}{2 \hbar^2 C_2}
    \Delta p^2 + o(\Delta p^2),
\end{equation}
implying that $p_x$ is the average value of the momentum, while its
typical fluctuation is given by
\begin{equation} \label{dp2}
    (\delta p_x)^2 = \hbar^2 C_2,
\end{equation}
and likewise for $y$ and $z$. Although Eqs.~(\ref{Gp}) and (\ref{dp2})
are consistent with Eqs.~(\ref{px}) and (\ref{C2b}), we note that
$\xi(\bp) = |\tilde \psi_A(\bp)|^2$ only if the integrals in
Eqs.~(\ref{psiAwp}) and (\ref{C1A}) are determined by the stationary
values of the integrands and the stationary values are unique. The
cumulants $C_1$ and $C_2$ can be regarded as functions of either $x$
or $p_x$, whenever a piece-wise bijective conformal mapping $p_x
\leftrightarrow x$ exists.

We have thus formally established generating functions for the
distributions of $\br$ and $\bp$, subject to caveats. Moreover, we
have established that the kinetic energy in the Schr\"odinger equation
can be presented as an average
\begin{align} E - V & = 
\left\langle \frac{(\langle \bp \rangle + \Delta \bp)^2}{2m}
\right\rangle \label{fl1} \\ &= \frac{\langle \bp \rangle^2}{2m} +
\frac{\langle \Delta \bp \rangle^2}{2m} \\ &= \frac{\bp^2}{2m} +
\frac{\hbar \nabla \bp}{2im} \label{fl3}
\end{align}
and that the terms $\bp^2/2m$ and $\hbar \nabla \bp/2im$ in the
r.h.s. of Eq.~(\ref{Riccati}) can be associated with the momentum
current and its fluctuations, respectively, since $\bp =
\bJ/|\psi|^2$. As a simple illustration, consider the ground state of
the one-dimensional Harmonic Oscillator, $V = m \omega^2 x^2/2$. Since
$p_x = i m \omega x$, one has $p_x^2/2m + V = 0$ throughout and,
consequently, $E = \hbar p_x'/2im = \hbar \omega/2$. In other words,
the zero-point energy is exclusively due to fluctuations of the
complex momentum.

That the last term in Eq.~(\ref{Riccati})---or
Eq.~(\ref{fl3})---accounts for the fluctuations of the momentum is
consistent with its quantum origin discussed following
Eq.~(\ref{Newton}). One may expect, then, that this term is also
responsible for phase shifts in the wavefunction, if any, that may
result during scattering. To see this explicitly, we consider
one-dimensional motion and substitute $p_\text{cl}^2/2m = E - V$ in
Eq.~(\ref{Riccati}):
\begin{equation} \label{ppcl}
  p - p_\text{cl}  = \frac{i \hbar p'}{p + p_\text{cl}}.  
\end{equation}
This yields
\begin{align} \label{phgainFull}
  \frac{1}{\hbar }\int^x \left[p(\tilde x) - p_\text{cl}(\tilde x)
    \right] d \tilde x &= i \int^x \!\!\! \frac{p' d\tilde x}{p +
    p_\text{cl}} \\ &= i \int^{p(x)} \hspace{-2mm} \frac{dp(\tilde
    x)}{p(\tilde x) + p_\text{cl}(\tilde x)}, \label{phgainFull2}
\end{align}
where we explicitly indicate that the last integral is parametric,
$\tilde x$ being the parameter that specifies the integration contour;
$p_\text{cl}(\tilde x)$ is not a constant. The real part of the
r.h.s. of the Eq.~(\ref{phgainFull}) or Eq.~(\ref{phgainFull2}) thus
yields the phase gain due to the quantum effects.

Consider first the Harmonic Oscillator and make (\ref{phgainFull2}) a
closed-loop integral counterclockwise around the branch cut connecting
the two turning points along the real axis. The Harmonic Oscillator is
special because its classical momentum $p_\text{cl}$ has no other
singularities in the complex plane, while its quantum momentum $p$ has
no singularities other than the poles that are all confined to the
classically-allowed portion of the real axis. Thus one may expand the
integration loop in Eq. ~(\ref{phgainFull2}) to become so large as to
make $p$ arbitrarily close to $p_\text{cl}$ at every point on the
loop, whereby $p = p_\text{cl} + O(1/x)$, according to
Eq.~(\ref{ppcl}).  Consequently the integral in
Eq.~(\ref{phgainFull2}) is equal to $\oint d p/2 p = (1/2) \oint d \ln
p = \pi i$. In view of Eqs.~(\ref{qc0}) and (\ref{BSqcont}), we
establish that the Bohr-Sommerfeld condition must be exact for the
Harmonic Oscillator.

When a large-$x$ expansion is not practical, one may, instead, attempt
an expansion in terms of $\hbar$. Eqs.~(\ref{phgainFull}) and
(\ref{phgainFull2}) indicate that to obtain a correction to the
classical value of $S$ in the lowest non-trivial order in $\hbar$, one
may simply replace $p$ by $p_\text{cl}$ in Eq.~(\ref{phgainFull2}):
\begin{equation} \label{phgain0}
  \frac{1}{\hbar }\int^x \left[p(\tilde x) - p_\text{cl}(\tilde x)
    \right] d \tilde x = \frac{i}{2} \int^{p(x)} \frac{d \tilde p}{\tilde p},
\end{equation}
Thus in the lowest order in $\hbar$, the phase gain caused by
reflection is $(-\pi/2)$, since $\ln (-1) = i \pi$. Scattering
produces no phase gain at this accuracy. Hereby we have reproduced the
standard WKB results, but without having to deal with the Van Vleck
determinant.~\cite{doi:10.1073/pnas.14.2.178, PhysRev.48.549,
  10.1063/1.1705112, BerryMount1972, 10.1063/1.470898, maslov2001semi}


Alternatively, one may substitute $p_\text{cl}^2/2m = E - V$ in
Eq.~(\ref{RiccatiSC}). Again, working in the lowest order in $(p -
p_\text{cl}) \propto \hbar$, yields
\begin{equation} 
  p - p_\text{cl} = - i \hbar \frac{m V'}{2 p_\text{cl}^2} + o(\hbar)
  = - \frac{i \hbar}{4} \, \frac{V'}{E - V} + o(\hbar).
\end{equation}
In turn, this implies
\begin{equation} \label{phgain}
  \frac{1}{\hbar }\int^x \left[p(\tilde x) - p_\text{cl}(\tilde x)
    \right] d \tilde x = \frac{i}{4} \int^{V(x)} \frac{dV}{V - E},
\end{equation}
up to corrections of higher order in $\hbar$.  The integral on the
r.h.s. of Eq.~(\ref{phgain}) does not depend on the explicit form of
the potential. The integral is relatively simple because, unlike the
classical momentum $p_\text{cl}$, the potential $V$ is a single-valued
function in the whole complex plane, no branch cuts needed. According
to Eq.~(\ref{phgain}), the phase gain in the semiclassical limit is
simply $(-\pi/2)$ times an integer number. The latter integer number
is sometimes referred to as the Maslov index.~\cite{10.1063/1.470898,
  maslov2001semi}

\section{Summary and final remarks}
\label{conclusions}

We have developed a mathematical representation of solutions of the
non-relativistic Schr\"odinger equation, by associating the spatial
variation of the wavefunction in the complex plane with the spatial
distribution of a momentum-like quantity $\bp$, a quantum mechanical
analog of the classical kinematic momentum. The momentum $\bp$ is the
complex flux normalized by the density and thus corresponds to
translation of inertial matter in the complex plane. The complex flux
is non-vanishing in most of the plane even if there is no net particle
flux along the real axis. For a bound state, the streamlines are
characteristic of a rigid-body rotation at infinity; the rotation
itself is inherent and underlies the zero-point vibrations. During
scattering, some---and sometimes all---of the streamlines of the
incoming signal are diverted away from the real axis toward the
imaginary axis.

The canonical component $\bP^*$ of the complex conjugate $\bp^*$
reflects the spatial variation of the wavefunction's phase. The
streamlines of the complex conjugate momenta comprise an
incompressible fluid. This conservation law underlies the invariance
of the Schr\"odinger equation with respect to analytic continuation
into the complex plane, as well as the invariance of certain
closed-loop integrals, in the complex plane, corresponding to
adiabatic invariants in classical mechanics. These closed-loop
integrals change, if at all, in discrete increments of fixed
magnitude. Each such increment corresponds to an excitation by exactly
one quantum, on the one hand, and with the appearance of a node in the
wave function, on the other hand.

The flows of the kinematic momentum are smooth within extended
segments of the complex plane, where they can be well approximated by
analytically continued classical momentum. The segments are separated
by linear arrays of simple poles, each pole centered at a zero of the
wavefunction. This singular behavior---by way of a diverging momentum
at the poles---may appear to imply a superliminal behavior. In turn,
this would seem to suggest that the poles are artifacts of the
non-relativistic limit $c \to \infty$. This is not the case, however,
since each pole's residue is equal to $\hbar/i$ and does not involve
the particle mass or the forces due to the potential. (By a similar
token, we do not attribute divergent velocities to a particle's
angular momentum.) For comparison, artifacts of the non-relativistic
limit do appear at large $x$, as the particle accelerates owing to the
potential forces: $p \sim [(2m)(E-V)]^{1/2}$. In the latter case,
inertial effects are explicitly involved.

The geometric, non-inertial nature of the momentum poles can be viewed
as an emergent behavior. For a stationary quantum state at energy $E$,
one may present the kinetic energy as a sum of two parts, due to the
steady part and to the fluctuating part of the momentum,
respectively. This is seen most readily when the vector potential is
vanishing, in which case $\bp = (\hbar/i) \nabla \psi$,
c.~f. Eqs.~(\ref{fl1})-(\ref{fl3}):
\begin{align}
  E - V & = \frac{1}{2m} \left\{ \bp^2 + \left[ (\hbar/i) \nabla \bp
    \right] \right\} \label{Ric1} = \frac{1}{2m} \left\{ \bp^2 +
  \left[ \hat \bP \bp \right] \right\} \\ & = \frac{1}{2m}
  \left\{ \langle \bp \rangle^2 + \left[ \langle \bp^2 \rangle -
    \langle \bp \rangle^2 \right] \right\} \\ & = -\frac{\hbar^2}{2m}
  \left\{ \left(\frac{\nabla \psi}{\psi} \right)^2 +
  \left[\frac{\nabla^2 \psi}{\psi} -\left(\frac{\nabla \psi}{\psi}
    \right)^2 \right]\right\} \\ &= -\frac{\hbar^2}{2m} \frac{\nabla^2
    \psi}{\psi}. \label{Ric4}
\end{align}
Here the square brackets are used to delineate the contribution of the
fluctuation throughout. Eq.~(\ref{Ric4}) yields the Schr\"odinger
equation, $(E - V) \psi = - (\hbar^2/2m) \nabla^2 \psi $, but only if
we supplement it by imposing an additional constraint that
\begin{equation} \label{bc0}
  \nabla^2 \psi(x_0) = 0, \text{ if } \psi(x_0) = 0.
\end{equation}
Without this additional constraint, Eq.~(\ref{Ric4}) is indeterminate
when $\psi=0$ strictly. Incidentally, inclusion of the fluctuation
part in Eq.~(\ref{Ric1}) can be seen as a necessary condition for the
Schr\"odinger equation to be linear in $\psi$.

If, instead, we insisted that $\bp$ in Eq.~(\ref{Ric1}) be well
behaved throughout, this would prevent us from having wavefunctions
that exhibit zeros, as well as the corresponding energy
values. Consequently, if one were to start from the description in
Eq.~(\ref{Ric1}), one must accept that there are putative isolated
points---or isolated defects in the complex plane, if you will---where
Eq.~(\ref{Ric1}) does not apply. Eq.~(\ref{bc0}) then serves as a
boundary condition that specifies the behavior of the function at
those putative defect locations.  Whether the poles are present, in
the first place, and what their locations $x_0$ are must be determined
self-consistently---alongside the energy $E$---which is equivalent to
saying that the poles represent an emergent behavior. In practice, one
may accomplish this self-consistent determination by solving the
stationary Schr\"odinger equation for its eigenfunctions and the
corresponding eigenvalues, subject to boundary conditions of interest,
but other methods are conceivable.

One way to drive home that the poles are emergent and result from an
instability is to imagine setting up the task of solving
Eq.~(\ref{Ric1}) or evaluating the wavefunction as a numerical
calculation. When solving Eq.~(\ref{Ric1}) numerically using finite
differences, the axis of the vortex cannot fall on a grid node since
the momentum must be finite at each node by construction; hence no
poles will be found. On the other hand, one routinely computes
solutions of the Schr\"odinger equation using power-law
expansions.~\cite{AS} There are always two solutions in classically
forbidden regions, one is exponentially decaying and normalizable, the
other exponentially increasing and non-normalizable. We are interested
in the former, more often than not. When consecutive expansion terms
are computed iteratively---which is typically the case---a numerical
instability may occur whereby the exponentially increasing counterpart
appears, too. At the same time, we have seen in Fig.~\ref{HOunNorm}
and in Appendix~\ref{rays} that a sector hosting the exponentially
diverging term must be flanked by strings of momentum poles.

Yet another perspective on standing waves being a symmetry-lowered
state that had resulted from an instability is afforded by recalling
that the Schr\"odinger equation along the real axis is, nominally, the
result of optimization of the functional~\cite{LLquantum}
\begin{equation}
  \frac{\delta}{\delta \psi} \int \left[ \frac{\hbar^2}{2m} |\nabla_1
  \psi|^2 + \left(V - E \right) |\psi|^2 \right] d^3 \br_1 = 0.
\end{equation}
This functional is unstable toward the formation of oscillating
solutions $\psi \propto e^{i \bk \br_1}$ for real-valued $k <
\sqrt{2m(E-V)}/\hbar$ in classically allowed regions, where $V < E$. A
standing wave is formed by linearly combining two such solutions,
$\psi_1$ and $\psi_2$, that are degenerate and mutually
linearly-independent: $\psi = \alpha \psi_1 + \beta \psi_2$. This
notion applies to motion along any anti-Stokes line and is a simple
way to rationalize the spacing between adjacent nodes of the
wavefunction. The $\beta/\alpha$ ratio is fixed by the potential $V$
and the boundary conditions. Although fixing the ratio breaks a
continuous symmetry, no Goldstone modes result; this is analogous to
what happens during a particular type of metal-insulator
transition.~\cite{LKgap}

The momentum poles and the accompanying critical points afford a vivid
visual representation of momentum flows in the complex plane, in the
form of fluid-like vortexes.  The divergence of the momentum at the
axis of a vortex is accompanied by a decrease in the wavefunction, to
yield a finite mass flux. Still, if one were to view the coordinate
and momentum as distributed quantities---as we did here---the
coordinate becomes sharply defined at each such pole, consonant with
standing waves exhibiting the remarkable phenomenon of quantum
entanglement. This is mirrored by momentum fluctuations becoming
divergent at each pole.

The classical and quantum momenta are characterized by rather
different types of singularities in the complex plane: branch cuts and
strings of poles, respectively. For the bound states of the Harmonic
Oscillator, the respective locations of these two sets of
singularities essentially mirror each other, while neither the
classical nor the quantum momentum exhibit singularities away from the
classically-allowed part of the real axis. For this reason, we have
found, the Harmonic Oscillator must satisfy the Bohr-Sommerfeld
quantization condition exactly.

A string of vortex-like objects in the complex plane is reminiscent of
a cross-section of a vortex sheet;~\cite{batchelor1967introduction}
such sheets commonly emerge in fluids as a result of Kelvin-Helmholtz
instabilities. At the same time, Eq.~(\ref{Newton}) has the same form
as the Navier-Stokes equation, even though the effective viscosity is
purely imaginary.  Yet the similarity with those fluid instabilities
is only limited: Consider a wavefunction $\psi(x_1, y_1)$ that has a
nodal line running along the $y_1$ axis in the (real) plane $(x_1,
y_1)$. Rotation in the plane $(x_1, x_2)$, at fixed $y_1$ and $y_2$,
corresponds to a translational motion of a planar object parallel to
the $(y_1, y_2)$ plane---even if along a closed loop---but not a
rotational motion around a line. By the same token, the vortexes in
the complex plane we considered here are dissimilar from those
underlying the vortex atom theory, due to Kelvin and
others.~\cite{KraghVortex}


{\bf Acknowledgments}: I thank Eric Bittner and Alexey Tcherniak for
useful conversations. This work was supported by a grant from the
Texas Center for Superconductivity at the University of Houston and,
in part, by the NSF Grant CHE-1956389.

\appendix

\section{Emergence of strings of poles,  wavefunction's normalization: A symmetry perspective}
\label{rays}

This Appendix presents a symmetry perspective on how, on the one hand,
strings of momentum poles come about that shape the complex currents
and, on the other hand, how bound solutions of the Schr\"odinger
equation arise. The two phenomena are intrinsically connected.

\begin{figure}
\centering
\includegraphics[width=  \columnwidth]{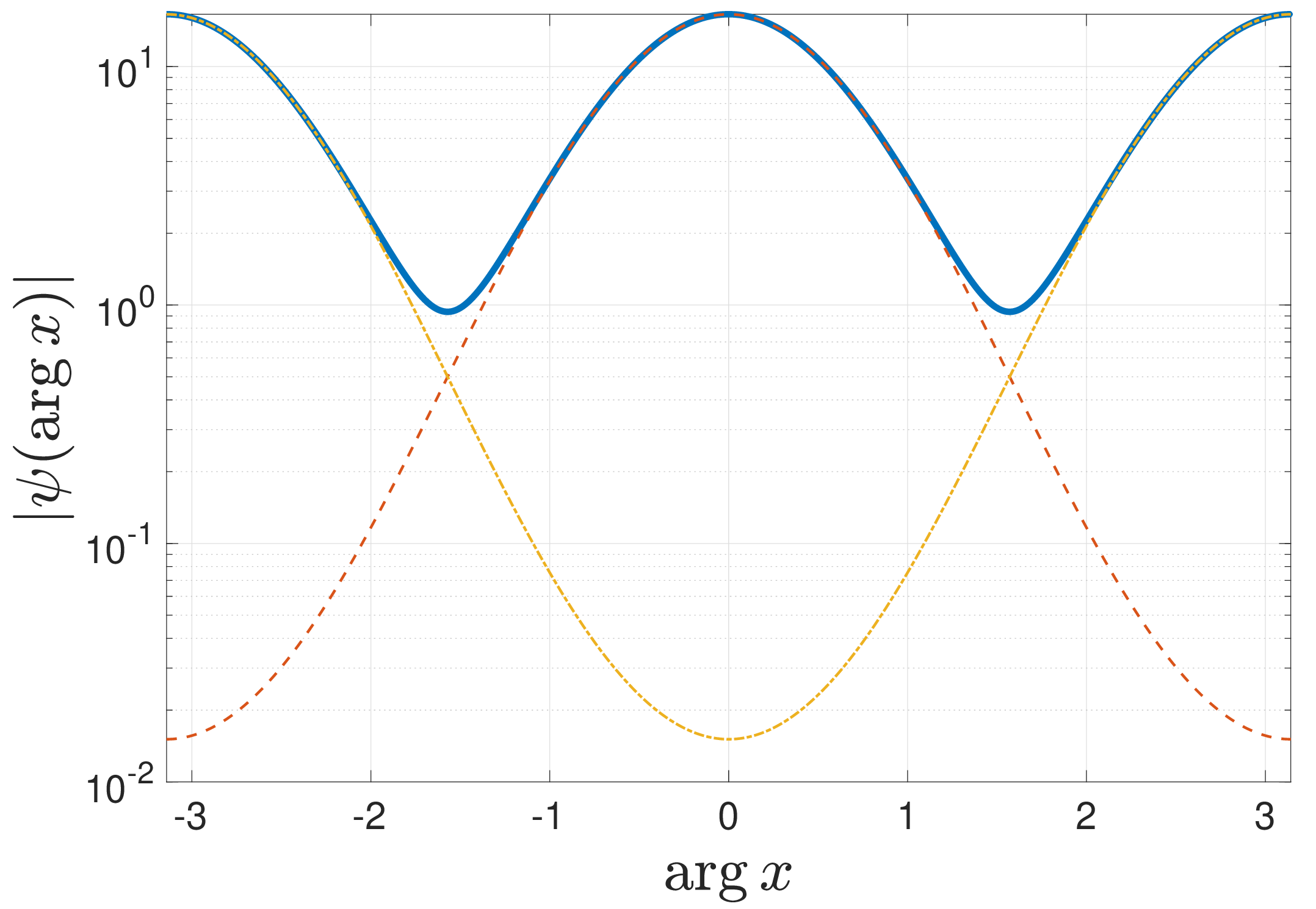}
\caption{Three solutions for the free particle at a negative energy $E
  = -1$, as functions of $\arg x$, at $|x| = 3.6$. The solid line is
  the trivial representation $\psi_0 = 2 \cosh x$ of the $C_2$ group of
  rotations in the complex plane. The dashed line is the basis
  function $\psi_\tb(x) = e^{x}$ centered in the Stokes sector $|\arg
  x| < \pi/2$, the dash-dotted line the ``basis'' function
  $\psi_\tb(x) = e^{-x}$ centered in the Stokes sector $|\arg x - \pi|
  < \pi/2$. $m = \hbar = 2$.} \label{expFig}
\end{figure}

To set the stage, consider a free particle in one spatial dimension at a negative energy $E<0$; this is directly relevant for motion in classically forbidden regions. The Schr\"odinger equation for a free particle is invariant with respect to rotation by the angle $\pi$ in the complex plane, because $\partial^2/\partial (x e^{i \pi})^2 = \partial^2/\partial x^2$. Let us set $\hbar = m = 2$ for convenience. Consider the pair of linearly independent solutions $\psi_0 = 2 \cosh x$ and $\psi_1 = 2 \sinh x$, which are, respectively, even and odd under the rotation:
\begin{align}
    \psi_0(x) &= \psi_\tb(x) + \psi_\tb(x e^{i \pi}) \\
    \psi_1(x) &= \psi_\tb(x) - \psi_\tb(x e^{i \pi})
\end{align}
and $\psi_\tb = e^{x}$ for the free particle case. We need not concern
ourselves with the overall normalization of our wavefunctions. The
label ``$\tb$'' alludes to the word ``basis'' since for large $x$,
$e^{x}$ is localized and sharply peaked, as a function of $\arg x$, in
the sector $|\arg x| < \pi/2$, while $\exp(x e^{i \pi}) = e^{-x}$ is
localized and sharply peaked in the complementary sector $|\arg x -
\pi| < \pi/2$, see Fig.~\ref{expFig}. The function pair $\psi_0$ and
$\psi_1$ is, then, quite analogous to a symmetry-adapted set of
molecular orbitals for a diatomic homonuclear molecule.~\cite{ABW}
Within this chemical analogy, the functions $\psi_\tb(x)$ and
$\psi_\tb(x e^{i \pi})$ together comprise an ``atomic'' basis set.

Despite its simplicity, the free particle case suffices to illustrate
most of the qualitative features of solutions of the Schr\"odinger
equation for polynomial potentials. Consider the even solution $\psi_0
= 2 \cosh(x) = e^x + e^{-x}$. It is exponentially large both in the
sector $|\arg x| < \pi/2$ and in the complementary sector $|\arg x -
\pi| < \pi/2$. However at the boundary between the two sectors, made
of the union of the rays $\arg x = 0$ and $\arg x = \pi$ respectively,
the function is purely oscillatory and, moreover, represents a
standing wave: $\psi_0(x) = 2 \cos(x_2)$. Thus, each of these rays
corresponds to a string of wavefunction's zeros and, consequently, of
momentum poles.

In the sector $|\arg x| < \pi/2$ , the component $e^{-x}$ of the
function $2 \cosh(x)$ is exponentially smaller than the dominant
component $e^x$, and vice versa in the sector $|\arg x - \pi| <
\pi/2$. Lines along which the dominant term is maximally dominant over
the subdominant term---$\arg x = 0$ and $\arg x = \pi$, respectively,
in the free particle case---are called Stokes
lines,~\cite{BerryMount1972} while the sectors themselves are called
Stokes sectors.~\cite{Steinmetz2014} As one moves sufficiently close
to the boundary of the sector, the dominant and subdominant components
approach each other in magnitude. Upon crossing the boundary, the
formerly subdominant term now becomes the dominant term and vice
versa, see Fig.~\ref{expFig}. Rays separating adjacent Stokes
sectors---$\arg x = + \pi/2$ and $\arg x = - \pi/2$, respectively, in
the free particle case---are referred to as anti-Stokes
lines~\cite{BerryMount1972} or Stokes rays.~\cite{Steinmetz2014} We
will avoid using the latter notation.

It so happens that the functions $\exp(x)$ and $\exp(x e^{i \pi})$ are
each a solution of the Schr\"odinger equation for the free particle,
too. Each of these two solutions is exceptional per the classification
in Ref.~\cite{Steinmetz2014}: An exceptional solution must decay
exponentially at least in one of the Stokes sectors. Because only one
of the exponentials is present inside the pertinent sector, the
wavefunction at the edges of the sector is still wave-like---$e^{\pm i
  x_2}$ in the free particle case---but it is no longer a standing
wave. Consequently, the anti-Stokes lines flanking the sector do not
house strings of poles. This is of direct relevance to bound states,
as we saw in Fig.~\ref{HOunNorm}.

Consider now a non-vanishing polynomial potential. The complex flows
at infinity are determined by the dominant term in the polynomial,
make $n$ its order. Consider a stationary solution at energy $E$. The
term $E \psi$ is sub-dominant at large $x$. The Schr\"odinger equation
becomes, to the leading order:
\begin{equation} \label{SchPower}
    -\frac{\partial^2 }{\partial x^2} \psi(x) + V_0 \, x^n \, \psi(x) = 0.
\end{equation}
This equation is invariant with respect to rotation in the complex
plane by an integer multiple of the angle $2 \pi/(n+2)$. Consistent
with this notion, there are $(n+2)$ Stokes sectors. For $V_0 > 0$, the
Stokes lines point along directions $\arg x = 2 l \pi /(n+2)$, $l = 0,
1, \ldots, (n+1)$. (These correspond to $\int^x V^{1/2}(\tilde x) d
\tilde x$ tending to become purely imaginary-valued for large $x$.)
For $V_0 < 0$, the Stokes lines are directed at $\arg x = (2l+1)
\pi/(n+2)$, $l = 0, 1, \ldots, (n+1)$. We can use the index $l$ to
label the sectors.

Consider $V_0 > 0$ for now, so that the potential is stable along the positive direction on the real axis. The dominant and subdominant asymptotic solutions within the Stokes sector $l=0$ are $\propto \exp[\pm x^{n/2+1}/(n/2+1)]/x^{n/4}$, per the WKB approximation.~\cite{LLquantum} Because of the rotational symmetry $C_{n+2}$ we just alluded to, one may seek solutions in the form:
\begin{equation} \label{Bloch}
    \psi_k(x)  = \sum_{l = 0}^{n+1} e^{i k \frac{2\pi l}{n+2}} \psi_\tb(x e^{-i \frac{2 \pi l}{n+2}}),
\end{equation}
By construction, the basis function $\psi_\tb(x e^{-i 2 \pi l/(n+2)})$
is sharply peaked, as a function of $\arg x$, inside the Stokes sector
$l$, $|\arg x - 2 \pi l/(n+2)| < \pi /(n+2)$, $l = 0, 1, \ldots,
(n+1)$. The peak is centered at the respective Stokes line and decays
away from the latter line all the way to the adjacent Stokes lines, to
become comparable to the subdominant exponential, after which it
continues to stay subdominant. In contrast with the free particle
case, $\psi_\tb(x e^{-i 2 \pi l/(n+2)})$ is no longer a solution of
the Schr\"odinger equation, but it is, nonetheless well approximated
by the dominant term in the asymptotic expansion within the Stokes
sector $l$.

The $\arg x$ dependence of the symmetry-adapted expansion
(\ref{Bloch}) can be thought of as a Bloch wave for a closed chain of
$(n+2)$ sites, in the tight-binding approximation;~\cite{ABW} the
reader will recognize the factors $e^{i k 2\pi l/(n+2)}$ as the
characters of the point group $C_{n+2}$. By construction, then, the
rotation $x \to x e^{i 2 \pi r/(n+2)}$ sends $\psi_k$ to itself times
the number $e^{i 2k \pi r/(n+2)}$, whose modulus is unity. Because the
representations $\psi_k$ are one-dimensional, one generically expects
solutions of Eq.~(\ref{SchPower}) to be just one of the functions
$\psi_k$. These each have an exponentially growing component in every
Stokes sector. Accordingly, Steinmetz calls solutions exhibiting an
exponentially growing term in each Stokes sector
generic.~\cite{Steinmetz2014} For generic solutions, then, each
anti-Stokes line hosts an infinite string of momentum poles, whose
azimuthal location depends only weakly, logarithmically, on the
prefactors of the two exponents; see Ref.~\cite{GUNDERSEN202294} for a
systematic discussion.

But general considerations mandate that the second order differential
equation (\ref{SchPower}) have two linearly-independent solutions,
implying that at least two of the $\psi_k$'s actually form a
degenerate pair. There must be, then, an underlying symmetry in the
problem other than the $C_{n+2}$ symmetry. And indeed, the equation
has real-valued coefficients; thus it can be analytically continued in
two equivalent ways: $x_1 \to x$ or $x_1 \to x^*$. This latter
symmetry guarantees that there will be always two functions $\psi_k$
that each solve the stationary Schr\"odinger equation---with a
potential that is real-valued on the real axis---and, at the same
time, could be linearly combined to form solutions that are even and
odd, respectively, under the transformation $x_2 \leftrightarrow -
x_2$. Now that one has at least two linearly independent $\psi_k$ at
their disposal, it is possible to make linear combinations giving rise
to an exceptional solution, i.e., such that at least one of the terms
$\psi_b(x e^{-i 2 \pi l/(n+2)})$ drops out, thus making the
wavefunction normalizable along the respective Stokes line.

\begin{figure}
\centering
\includegraphics[width= 0.8 \columnwidth]{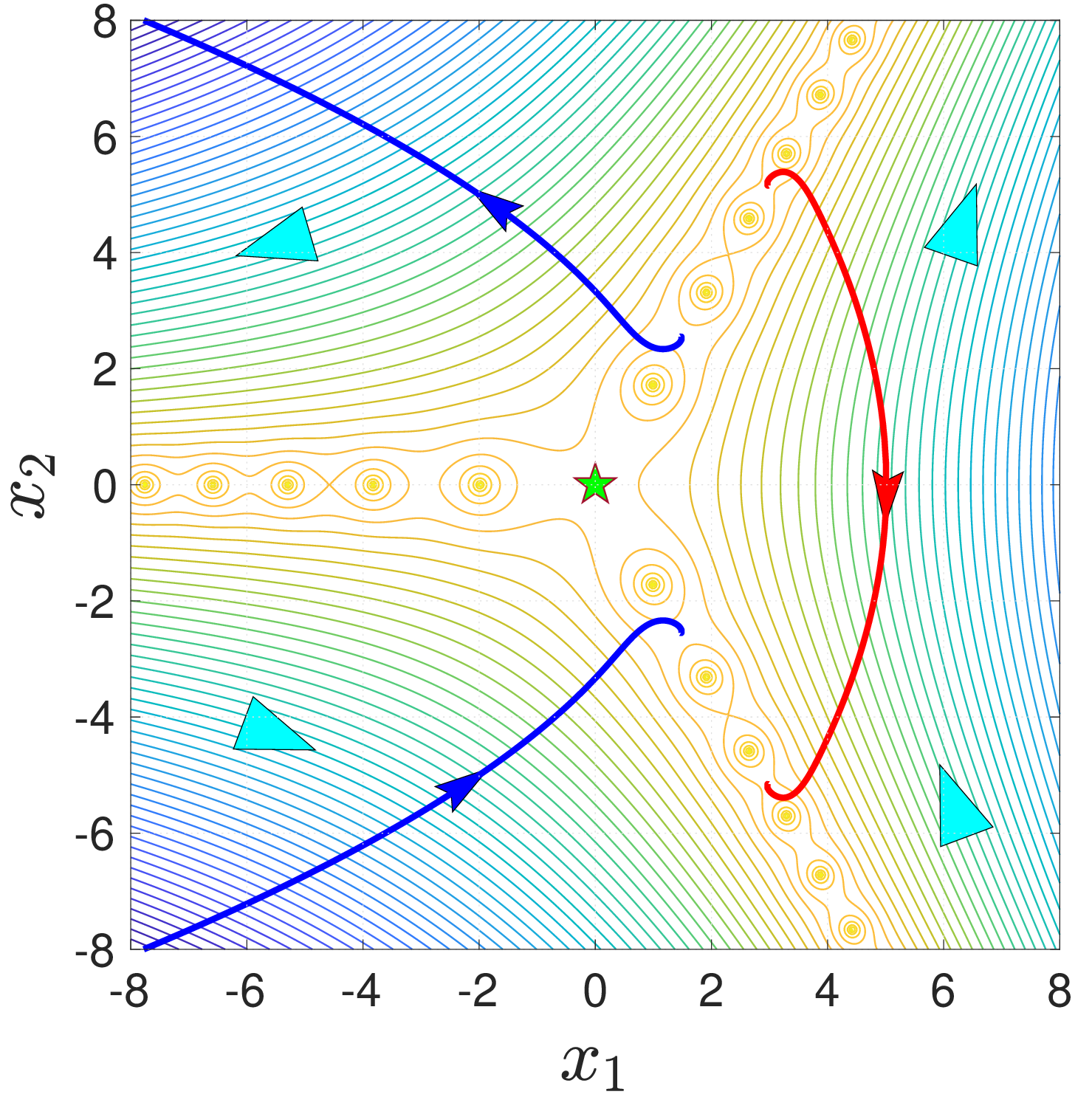}
\caption{The solution of the Airy equation in the form of the trivial
  representation $\psi_0(x) = \text{Bi}(x) + 3^{1/2} \text{Ai}(x)$ of
  the point group $C_3$ for rotations in the complex plane. Blue and
  red lines with arrows exemplify trajectories $dx/p = d\tau$,
  infinitesimal and real-valued. The star indicates the classical
  turning point.  The contour lines show lines of constant density
  $|\psi|^2$. The cyan triangles indicate direction of phase flows,
  see Section~\ref{continuum}. $m = \hbar = 2$.} \label{AiryFig}
\end{figure}

As an illustration, consider the case $n=1$, whereby
Eq.~(\ref{SchPower}) becomes the venerable Airy equation. One can use
Eqs.~(10.4.2) and (10.4.3) (or Eq.~(10.4.6)) of Ref.~\cite{AS} to see
that the trivial representation is $\psi_0(x) = \text{Bi}(x) + 3^{1/2}
\text{Ai}(x)$, where $\text{Ai}(x)$ and $\text{Bi}(x)$ are the Airy
functions of the first and second kind, respectively. The
corresponding complex currents are shown in Fig.~\ref{AiryFig}; the
three anti-Stokes lines are immediately identified by the three
strings of poles emanating from the origin. Using consistent
normalization, one can further show that the solution normalizable
along (the positive ray of) the real axis is given by $\text{Ai} =
(\psi_0 - \psi_1)/2 \sqrt{3}$, while the divergent counterpart is
$\text{Bi} = (\psi_0 + \psi_1)/2$. $\text{Ai}(x)$ is thus an
exceptional solution that has only one ray of momentum poles---running
along the negative half of the axis---which represents the
interference pattern between the incident and reflected wave,
respectively, the classical turning point being at $x=0$. On a side
note, knowing $\psi_0$ and $\psi_1$ is insufficient to readily
determine the basis function $\psi_\tb$, but a decent asymptotic
approximation that covers $-2 \pi/3 < \arg(x) < 2 \pi/3$ can be
obtained by neglecting, in Eq.~(\ref{Bloch}), $\psi_\tb(x e^{i 2
  \pi/3})$ for $0 < \arg x < 2 \pi/3$ and $\psi_\tb(x e^{-i 2 \pi/3})$
for $- 2 \pi/3 < \arg x < 0$. One thus obtains, after some algebra,
simple expressions $\psi_\tb(x) \sim \text{Bi}(x) - i \text{Ai}(x)$
for $0 < \arg x < 2 \pi/3$, and $\psi_\tb(x) \sim \text{Bi}(x) + i
\text{Ai}(x)$ for $- 2 \pi/3 < \arg x < 0$. This approximate form
becomes inaccurate close to the Stokes lines $\arg x = -2 \pi/3, 0, +
2 \pi/3$, the error being small in the sense that it is comparable in
magnitude to the subdominant exponential term. For instance, this
approximate term exhibits a small discontinuity, $2 i \text{Ai}(x_1)$,
at $\arg x = 0$. This is a manifestation of the so called Stokes
phenomenon.~\cite{10.1098/rspa.1989.0018, Berry1988, ParisWood,
  MeyerStokes}

The $n=2$ case corresponds to a bound state for $V_0 >0$ and scattering/tunneling for $V_0 <0$. Let us write out the four symmetry adapted solutions of Eq.~(\ref{SchPower}) explicitly:
\begin{align}
    \psi_0 &= \psi_\tb(x) + \psi_\tb(x/i) + \psi_\tb(-x) + \psi_\tb(i x) \\
    \psi_1 &= \psi_\tb(x) + i \psi_\tb(x/i) - \psi_\tb(-x) - i \psi_\tb(i x) \\
    \psi_2 &= \psi_\tb(x) - \psi_\tb(x/i) + \psi_\tb(-x) -  \psi_\tb(i x) \\
    \psi_3 &= \psi_\tb(x) - i \psi_\tb(x/i) - \psi_\tb(-x) + i \psi_\tb(i x) 
\end{align}
As in the $n=3$ case, it is always possible to construct a solution
that is normalizable at least on one end of the real axis. The
parabolic cylinder function $U(a, x)$~\cite{AS} accomplishes just
that; specifically the wavefunction decays in the positive direction,
by construction.

If, on the other hand, one were to form solutions for Eq.~(\ref{SchPower}) that are normalizable on both ends of the real axis at the same time, those solutions would have to involve special combinations of the $\psi_k$'s. Indeed, the even bound states of the harmonic oscillator only result from the combination $(\psi_0 - \psi_2)$, while the odd bound states only result from the combination $(\psi_1 - \psi_3)$. 

The bound states are of lower symmetry than the equation itself, as
caused by our imposing the boundary conditions. In geometric terms,
this symmetry lowering presents as the momentum currents $p$ having
the same orientation everywhere at infinity. The symmetry-lowered set
has the $C_2$ symmetry. The non-trivial representation of this group
transforms equivalently to an eigenstate of an odd-numbered,
integer-valued angular momentum. In addition, both representations
each have two nodes located at the Stokes lines $l=0$ and $l=2$
respectively: $0 \cdot \psi_\tb(x) + 1 \cdot \psi_\tb(x/i) + 0 \cdot
\psi_\tb(-x) \pm 1 \cdot \psi_\tb(i x)$. Because this corresponds to a
standing wave that fits exactly one period over the full rotation by
$2 \pi$---as in $\cos \varphi = (e^{i 1 \cdot \varphi} + e^{-i 1 \cdot
  \varphi})/2$---we conclude that rigid body-like flows accompanying
bound states of the harmonic oscillator at infinity correspond to a
quantized angular momentum of magnitude $1 \cdot \hbar$.

In the $V_0 < 0$ case, the anti-Stokes lines are pointed along the
directions $\arg x = l \pi/2$, $l = 0, 1, 2, 3$. To ensure that
$\psi_{x_1 \to + \infty}$ consists solely of a transmitted wave, there
should be no poles in the vicinity of the ray $\arg x = 0$. Since the
wave is outgoing, $\Re p > 0$ on the real axis at large $x$, the
imaginary component of the momentum $p \propto + \sqrt{x^2 - E/V_0}$
must be positive in the first quadrant of the complex plane, too. We
conclude, then, that the Stokes sector $0 < \arg x < \pi/2$ houses
only the decaying term. In turn, this implies the ray $\arg x = \pi/2$
does not host poles either, while the second and fourth quadrants must
house exponentially growing terms. The negative real axis, left of the
classical turning point, must have a string of poles nearby---because
both incident and reflected signals are present---while the string is
shifted upwards so that the net mass flux at the real axis is
positive. (The momentum $p \propto + \sqrt{x^2 - E/V_0}$ is negative
in the third quadrant of the complex plane, implying the vorticity is
positive.) Consequently, the third quadrant also has an exponentially
growing term, while there is a string of poles just off, to the right
of, the negative imaginary axis.

We reiterate that everywhere above, we referred exclusively to
large-$x$ behaviors of anti-Stokes lines. Close to the origin and for
higher-order polynomial potentials, the lines can form complicated
structures,~\cite{10.1063/1.3598419, Giller_2008} but analysis can be
aided by symmetry-based considerations.~\cite{AIF_2008__58_2_603_0,
  Eremenko2008} In any event, a few simple rules may be formulated: A
string of poles, if any, must run along an anti-Stokes line. Such a
line ends either close to a classical turning point or at infinity and
can, in fact, cross the whole plane
uninterrupted.~\cite{10.1063/1.3598419, Giller_2008,
  AIF_2008__58_2_603_0, Eremenko2008} If an anti-Stokes line hosts a
string of poles, it must be flanked by opposing $p^\ph$ currents. The
currents smooth out exponentially quickly, as one moves off the
anti-Stokes line in question. The respective rate is determined by the
local spacing between adjacent poles. Indeed, locally the variation of
the wave function is $\propto e^{i k x}$, where $k \propto p^*(x)$ is
the local direction of the anti-Stokes line. As one moves off the
anti-Stokes line, the momentum approaches its classical value
exponentially quickly, by the same token as in the free particle case
considered in the beginning of Section~\ref{flows}. In the absence of
poles along an anti-Stokes line, the streamlines of the phase momentum
are mutually aligned while remaining parallel to the anti-Stokes line.

\section{Calculation of wavefunctions}
\label{details}

The wavefunctions for bound states of the Harmonic Oscillator $V = m
\omega x^2/2$ were computed using the parabolic cylinder
function~\cite{AS} $U(a, x)$.  Solutions for scattering in the
inverted parabolic potential $V = - m \omega x^2/2$ are given by the
parabolic cylinder function~\cite{AS} $W(a, x)$. The parameters are
chosen $\hbar = m = 1/\omega = 2$, as in Ref.~\cite{BARTON1986322}, so
that the Schr\"odinger equation has the standard form of the Weber
equation, Chapter 19 of Ref.~\cite{AS}. The computation itself was
performed using MATLAB functions created by
E. Cojocaru.~\cite{Cojocaru2026}

The solution for the inverted parabolic potential at $E=0$ was
evaluated according to
\begin{align}
  U(0, x) &= \left( \frac{x}{2 \pi} \right)^{1/2} K_{1/4} (x^2/4)
  \\ &= \left( \frac{x}{2 \pi} \right)^{1/2} \left( \frac{\pi}{2}
  \right) \frac{I_{-1/4} (x^2/4) - I_{1/4} (x^2/4)}{\sin(\pi/4)} \\ &=
  \frac{\pi^{1/2}}{2} \left[ 8^{1/4} \sum_{m=0} \frac{(x^2/8)^{2m}}{m!
      \, \Gamma(m+3/4)} \right. \nonumber \\ &- \left. 8^{-1/4}
    \sum_{m=0} \frac{x (x^2/8)^{2m}}{m! \, \Gamma(m+5/4)} \right]
\end{align}
and we used Eqs.~(19.15.9), (9.6.2), and (9.6.10) of Abramowitz and
Stegun.~\cite{AS} $K$ and $I$ are the modified Bessel functions.

The Airy functions were evaluated using MATLAB's built-in functions.

\bibliography{lowT}

\end{document}